\documentclass[aps,twocolumn,nofootinbib,preprintnumbers,superscriptaddress,floatfix]{revtex4-2}

\usepackage{color}
\usepackage{bbm}
\usepackage{tikz}
\usepackage{braket}
\usepackage{comment}
\usepackage[caption=false]{subfig}

\usepackage[
pagebackref=false,
colorlinks=true,
linkcolor=blue,
urlcolor=blue,
filecolor=black,
citecolor=red,
pdfstartview=FitV,
pdftitle={},
pdfauthor={},
pdfsubject={},
pdfkeywords={},
pdfpagemode=None,
bookmarksopen=true
]{hyperref}

\usepackage[normalem]{ulem}
\usepackage{amsmath, amssymb, bm}
\usepackage{enumerate}
\usepackage{amsfonts}
\usepackage{epsfig}
\usepackage{mathbbol}
\usepackage{mathrsfs}
\usepackage{amsthm}

\usetikzlibrary{shapes}
\usetikzlibrary{decorations.pathmorphing}
\tikzset{snake it/.style={decorate, decoration=snake}}

\def\s{\sigma}

\def\vbm{\vec{\mathbf{m}}}

\begin{document}
\title{Neural network enhanced cross entropy benchmark for monitored circuits}
\author{Yangrui Hu}
\email{yangrui.hu1@uwaterloo.ca}
\affiliation{Department of Physics and Astronomy, University of Waterloo, Ontario, N2L 3G1, Canada}
\author{Yi Hong Teoh}
\affiliation{Department of Physics and Astronomy, University of Waterloo, Ontario, N2L 3G1, Canada}
\author{William Witczak-Krempa}
\affiliation{D\'epartement de Physique, Universit\'e de Montreal, Montreal, QC H3C 3J7, Canada}
\affiliation{Centre de Recherches Math\'ematiques, Universit\'e de Montr\'eal, Montr\'eal, QC H3C 3J7, Canada}
\affiliation{Institut Courtois, Universit\'e de Montr\'eal, Montr\'eal, QC H2V 0B3, Canada}
\author{Roger G. Melko}
\affiliation{Department of Physics and Astronomy, University of Waterloo, Ontario, N2L 3G1, Canada}
\affiliation{Perimeter Institute for Theoretical Physics, Waterloo, Ontario, N2L 2Y5, Canada}

\begin{abstract}

We explore the interplay of quantum computing and machine learning to advance experimental protocols for observing measurement-induced phase transitions (MIPT) in quantum devices. 
In particular, we focus on trapped ion monitored circuits and apply the cross entropy benchmark recently introduced by [Li et al., Phys. Rev. Lett. {\bf 130}, 220404 (2023)], which can mitigate the post-selection problem. 
By doing so, we reduce the number of projective measurements\textemdash the sample complexity\textemdash required per random circuit realization, which is
a critical limiting resource in real devices.
Since these projective measurement outcomes form a classical probability distribution, they are suitable for learning with a standard machine learning generative model.
In this paper, we use a recurrent neural network (RNN) to learn a representation of the measurement record for a native trapped-ion MIPT, and show that using this generative model can substantially reduce the number of measurements required to accurately estimate the cross entropy.
This illustrates the potential of combining quantum computing and machine learning to overcome practical challenges in realizing quantum experiments. 

\end{abstract}

\maketitle

\section{Introduction}

Generative models have demonstrated unprecedented abilities in scaling~\cite{Kaplan:2020trn} and emergence~\cite{wei2022emergent}. Modern generative strategies, such as autoregressive language models, are capable of learning complex probability distributions given any number of diverse data sets.
Their ability to scale and generalize makes them well-suited for applications in quantum computing design, characterization, and control~\cite{carrasquilla2019reconstructing,torlai2020machine,sweke2020reinforcement,durrer2020automated,moon2020machine,teoh2020machine,carrasquilla2020machine,czischek2021miniaturizing,Dawid:2022fga,torlai2023quantum,zwolak2023colloquium,Melko2024-ev}. 
Recent quantum state preparation protocols, which utilize non-unitary circuits, represent the state-of-the-art in the field.  
However, these protocols require knowledge of probability distributions~\cite{Li:2022vks} of an exponentially large number of measurement outcomes, presenting a significant challenge. 
Despite these challenges, this field, which is often referred to as monitored circuits, poses a rich playground for the application of generative machine learning to influence the field of quantum computing in the near future.

Monitored circuits offer an opportunity to explore new physics, including phases and phase transitions~\cite{Skinner:2018tjl,PhysRevB.98.205136,PhysRevB.99.224307,Choi:2019nhg,Jian:2019mny,Li_2019,Zabalo:2019sfl,Bao_2020,PhysRevX.10.041020,PhysRevB.101.104302,Agrawal:2021ukw,Li_2021v2,Sang_2021,PhysRevLett.129.120604,Weinstein_2022,Skinner:2023hle,Li_KPZ2023,PhysRevB.108.054307,avakian2024longrangemultipartiteentanglementnear,PhysRevB.110.195107}, in systems that are not typically found in nature: a combination of quantum computers and (classical) observers~\cite{PhysRevB.100.134306,Fisher:2022qey,Agrawal:2023kuy}. 
Trapped ions have become a key experimental testing ground for the feasibility of monitored circuit experiments, thanks to the precise and programmable control achievable over individual qubits in a trapped-ion quantum computer~\cite{PhysRevLett.92.207901,PhysRevA.72.063407,PhysRevLett.100.200502,PhysRevA.78.062331,blatt2012quantum,Debnath_2016,10.1063/1.5088164,Christensen:2020dik,RevModPhys.93.025001,PRXQuantum.2.020343,Pino:2020mku,Noel:2021hez,Agrawal:2023kuy,Foss-Feig:2024blk}. 
Notably, pure quantum states can be prepared consistently~\cite{Debnath_2016,RevModPhys.93.025001}
and single-qubit measurements with a constant rate are feasible through optical addressing~\cite{10.1063/1.4900754,Shih:2021fcs}.

However, there are still a number of experimental and theoretical challenges, including the post-selection problem for examining the measurement-induced phase transition (MIPT) on a quantum computer.\,\footnote{
With the post-selection problem, MIPT can only be observed in a quantum computer with small system sizes~\cite{Koh:2022ajm}. It has been studied extensively to address the post-selection problem via various approaches and perspectives~\cite{PhysRevLett.126.060501,PhysRevX.12.011045,PRXQuantum.2.040319,Fisher:2022qey,potter2022entanglement,Noel:2021hez,Agrawal:2023kuy}.
}
To address these challenges, machine learning emerges as a powerful tool. 
The intuition is simple and natural as follows. 
In general, since monitored circuits are made up of both a quantum component and a classical component, the latter is a ripe playground for exploration with machine learning.
Indeed, neural networks have been used to find decoders that map measurement outcomes in monitored circuits to the reference qubit configuration~\cite{Dehghani:2022aia}.
In this paper, we leverage neural networks to address the sample complexity issue in the cross entropy benchmark. Let us explain it in more detail.   

The cross entropy benchmark (XEB) was applied in an experimental protocol for observing MIPTs on quantum computers in~\cite{Li:2022vks}, where the post-selection problem is mitigated. As a price to be paid, classical simulations of the random circuits in their protocol are necessary and restrict the scalability. 
Indeed, in a generic setting, exponentially long classical simulations are required~\cite{Li:2022vks}.\,\footnote{Note that for Clifford random circuits initialized with stabilizer states, this protocol is scalable~\cite{Li:2022vks}.} 
Moreover, for non-Clifford circuits, computing the cross entropy necessitates a significantly large number of measurement runs per circuit. The number of measurement runs is also closely tied to the experiment cost, which provides another practical motivation to reduce it~\cite{Iouchtchenko:2022php}.

The aim of this paper is two-fold. 
First, we validate the XEB for measurement-induced phase transition in trapped ion hybrid circuits by classical simulations. Second, we propose a recurrent neural network enhanced protocol for verifying MIPTs on quantum computers and demonstrate its effectiveness in enhancing the benchmark. 
The rest of this work is organized in the following manner. Section~\ref{sec:circuit} provides a review of the hybrid circuit constructed using native gates to a trapped ion device and projective measurements. In Section~\ref{sec:XEB}, we introduce and validate the cross entropy benchmark for observing a measurement-induced phase transition in the trapped ion hybrid circuit.
Section~\ref{sec:RNN} is devoted to the recurrent neural network model and analyzing the sample complexity. To do so, for a given circuit, we study the number of measurement runs required to achieve a cross entropy estimation with a target accuracy. 
We conclude with a discussion in Section~\ref{sec:discussion}, followed by supplementary details on numerics and RNN models presented in the appendices.

\section{Trapped ion monitored circuit}\label{sec:circuit}

In this work, we consider the quantum device consisting of $L$ trapped atomic ions and random quantum gates native to the device as in~\cite{Czischek:2021nso}. The native two-qubit entangling gates include the so-called M{\o}lmer-S{\o}rensen (MS) gate~\cite{PRXQuantum.2.020343,PhysRevLett.117.060505} which takes the following form
\begin{equation}
    \begin{split}
        M_{j,k}(\theta) ~=&~ \cos\theta\,\mathbb{I}_j\otimes\mathbb{I}_k ~-~i\,\sin\theta\,X_j\otimes X_k  ~~,\\
        ~=&~ \begin{pmatrix}
        \cos\theta & 0 & 0 & -i\sin\theta\\
        0 & \cos\theta & -i\sin\theta & 0 \\
        0 & -i\sin\theta & \cos\theta & 0 \\
        -i\sin\theta & 0 & 0 & \cos\theta 
    \end{pmatrix}\,,
    \end{split}
    \label{equ:MSgate}
\end{equation}
where $j$ and $k$ denote the two qubits it acts on. The above MS gate is essentially a rotation around the $XX$ axis and can be generalized to rotations around a generic axis.  
Note that in conventional random quantum circuits exhibiting the measurement-induced phase transition, the two-qubit entangling gate is typically sampled from a specific ensemble. In contrast, although in our case randomness can be introduced by randomizing the rotation angle $\theta$ and the axis, fixing the MS gate while incorporating random single-qubit rotations is sufficient.\,\footnote{This sufficiency was demonstrated exclusively through numerical methods in~\cite{Czischek:2021nso}. 
However, \cite{Kong:2024dfn} developed a mathematical framework to study the efficiency of the random circuits. 
It provides a theoretical foundation for random circuit models such that the entangling gates are fixed and the randomness only comes from single-qubit gates. 
In particular, they considered a relevant circuit model called the ``ironed gadget", which consists of single-qubit Haar-random gates and a fixed 2-qubit unitary gate. They systematically studied the speed of convergence for various choices of the fixed 2-qubit gate. 
It would be interesting to apply this framework in the (trapped ion) hybrid circuit design and benchmarking.}

\begin{figure}[htp!]
    \centering
    \begin{tikzpicture}
    \draw[thick, ->] (-2.5,-0.8-0.2) -- (-2.5,-0.8+0.25);
    \draw[thick, ->] (-1.5,-0.8-0.2) -- (-1.5,-0.8+0.25);
    \draw[thick, ->] (-0.5,-0.8-0.2) -- (-0.5,-0.8+0.25);
    \draw[thick, ->] (0.5,-0.8-0.2) -- (0.5,-0.8+0.25);
    \draw[thick, ->] (1.5,-0.8-0.2) -- (1.5,-0.8+0.25);
    \draw[thick, ->] (2.5,-0.8-0.2) -- (2.5,-0.8+0.25);
    \draw[thick, ->] (3.5,-0.8-0.2) -- (3.5,-0.8+0.25);
    \filldraw[thick,fill=magenta!30!white] (-2.5,-0.8) circle (0.1);
    \filldraw[thick,fill=magenta!30!white] (-1.5,-0.8) circle (0.1);
    \filldraw[thick,fill=magenta!30!white] (-0.5,-0.8) circle (0.1);
    \filldraw[thick,fill=magenta!30!white] (0.5,-0.8) circle (0.1);
    \filldraw[thick,fill=magenta!30!white] (1.5,-0.8) circle (0.1);
    \filldraw[thick,fill=magenta!30!white] (2.5,-0.8) circle (0.1);
    \filldraw[thick,fill=magenta!30!white] (3.5,-0.8) circle (0.1);
    \draw[thick] (0.5,-0.22) -- (0.5,-0.5);
    \draw[thick] (-2.5,-0.22) -- (-2.5,-0.5);
    \draw[thick] (-1.5,-0.22) -- (-1.5,-0.5);
    \draw[thick] (-0.5,-0.22) -- (-0.5,-0.5);
    \draw[thick] (1.5,-0.22) -- (1.5,-0.5);
    \draw[thick] (2.5,-0.22) -- (2.5,-0.5);
    \draw[thick] (3.5,-0.22) -- (3.5,-0.5);
    \node[draw, shape=rectangle, minimum width=5em] (U1) at (0,0){$U$};
    \node[draw, shape=rectangle, minimum width=5em] (U1) at (-2,0){$U$};
    \node[draw, shape=rectangle, minimum width=5em] (U1) at (2,0){$U$};
    \draw (3.9,0.235) -- (3.1,0.235) -- (3.1,-0.235) -- (3.9,-0.235);
    \draw[thick] (-2.5,0.22) -- (-2.5,0.22+1.2);
    \draw[thick] (-1.5,0.22) -- (-1.5,0.95);
    \draw[thick] (-0.5,0.22) -- (-0.5,0.95);
    \draw[thick] (0.5,0.22) -- (0.5,0.95);
    \draw[thick] (1.5,0.22) -- (1.5,0.95);
    \draw[thick] (2.5,0.22) -- (2.5,0.95);
    \draw[thick] (3.5,0.22) -- (3.5,0.95);
    \filldraw[thick,fill=cyan!30!white] (-2.5,0.22+0.73/2) circle (0.1);
    \filldraw[thick,fill=cyan!30!white] (0.5,0.22+0.73/2) circle (0.1);
    \filldraw[thick,fill=cyan!30!white] (-0.5,0.22+0.73/2) circle (0.1);
    \filldraw[thick,fill=cyan!30!white] (1.5,0.22+0.73/2) circle (0.1);
    \filldraw[thick,fill=cyan!30!white] (3.5,0.22+0.73/2) circle (0.1);
    \node[draw, shape=rectangle, minimum width=5em] (U1) at (1,1.2){$U$};
    \node[draw, shape=rectangle, minimum width=5em] (U1) at (-1,1.2){$U$};
    \node[draw, shape=rectangle, minimum width=5em] (U1) at (3,1.2){$U$};
    \draw[thick] (0.5,0.22+1.2) -- (0.5,0.95+1.2);
    \draw[thick] (-2.5,0.22+1.2) -- (-2.5,0.95+1.2);
    \draw[thick] (-1.5,0.22+1.2) -- (-1.5,0.95+1.2);
    \draw[thick] (-0.5,0.22+1.2) -- (-0.5,0.95+1.2);
    \draw[thick] (1.5,0.22+1.2) -- (1.5,0.95+1.2);
    \draw[thick] (2.5,0.22+1.2) -- (2.5,0.95+1.2);
    \draw[thick] (3.5,0.22+1.2) -- (3.5,0.95+1.2);
    \filldraw[thick,fill=cyan!30!white] (-1.5,0.22+0.73/2+1.2) circle (0.1);
    \filldraw[thick,fill=cyan!30!white] (0.5,0.22+0.73/2+1.2) circle (0.1);
    \filldraw[thick,fill=cyan!30!white] (2.5,0.22+0.73/2+1.2) circle (0.1);
    \filldraw[thick,fill=cyan!30!white] (3.5,0.22+0.73/2+1.2) circle (0.1);
    \node[draw, shape=rectangle, minimum width=5em] (U1) at (0,2.4){$U$};
    \node[draw, shape=rectangle, minimum width=5em] (U1) at (-2,2.4){$U$};
    \node[draw, shape=rectangle, minimum width=5em] (U1) at (2,2.4){$U$};
    \draw[thick] (0.5,0.22+2.4) -- (0.5,0.95+2.4);
    \draw[thick] (-2.5,0.22+2.4) -- (-2.5,0.95+2.4);
    \draw[thick] (-1.5,0.22+2.4) -- (-1.5,0.95+2.4);
    \draw[thick] (-0.5,0.22+2.4) -- (-0.5,0.95+2.4);
    \draw[thick] (1.5,0.22+2.4) -- (1.5,0.95+2.4);
    \draw[thick] (2.5,0.22+2.4) -- (2.5,0.95+2.4);
    \draw[thick] (3.5,0.22+2.4) -- (3.5,0.95+2.4);
    \draw (3.9,0.235+2.4) -- (3.1,0.235+2.4) -- (3.1,-0.235+2.4) -- (3.9,-0.235+2.4);
    \filldraw[thick,fill=cyan!30!white] (-2.5,0.22+0.73/2+2.4) circle (0.1);
    \filldraw[thick,fill=cyan!30!white] (0.5,0.22+0.73/2+2.4) circle (0.1);
    \filldraw[thick,fill=cyan!30!white] (-0.5,0.22+0.73/2+2.4) circle (0.1);
    \filldraw[thick,fill=cyan!30!white] (2.5,0.22+0.73/2+2.4) circle (0.1);
    \path [draw=magenta, snake it]
    (-2.5,3.7) -- (-1,3.7) -- (1,3.7) -- (3.5,3.7);
    \draw[thick, ->] (-2.5-0.1,3.7-0.2) -- (-2.5+0.1,3.7+0.2);
    \draw[thick, ->] (-1.5+0.1,3.7-0.2) -- (-1.5-0.1,3.7+0.2);
    \draw[thick, <-] (-0.5-0.1,3.7-0.2) -- (-0.5+0.1,3.7+0.2);
    \draw[thick, ->] (0.5+0.1,3.7-0.2) -- (0.5-0.1,3.7+0.2);
    \draw[thick, ->] (1.5-0.1,3.7-0.2) -- (1.5+0.1,3.7+0.2);
    \draw[thick, <-] (2.5+0.1,3.7-0.2) -- (2.5-0.1,3.7+0.2);
    \draw[thick, ->] (3.5-0.1,3.7-0.2) -- (3.5+0.1,3.7+0.2);
    \filldraw[thick,fill=magenta!30!white] (-2.5,3.7) circle (0.1);
    \filldraw[thick,fill=magenta!30!white] (-1.5,3.7) circle (0.1);
    \filldraw[thick,fill=magenta!30!white] (-0.5,3.7) circle (0.1);
    \filldraw[thick,fill=magenta!30!white] (0.5,3.7) circle (0.1);
    \filldraw[thick,fill=magenta!30!white] (1.5,3.7) circle (0.1);
    \filldraw[thick,fill=magenta!30!white] (2.5,3.7) circle (0.1);
    \filldraw[thick,fill=magenta!30!white] (3.5,3.7) circle (0.1);
    \draw[thick,<->] (-3,-1.5) -- (4,-1.5) node[anchor=north west] {$L$};
    \draw[thick,->] (-3.5,-0.2) -- (-3.5,4) node[anchor=south east] {$T$};
    \node[draw, shape=rectangle, minimum width=5em] (U1) at (6-7,1.2-4){$U$};
    \draw[thick] (5.5-7,0.22+1.2-4) -- (5.5-7,0.95+1.2-0.73/2-4);
    \draw[thick] (6.5-7,0.22+1.2-4) -- (6.5-7,0.95+1.2-0.73/2-4);
    \draw[thick] (5.5-7,0.22+0.73/2-4) -- (5.5-7,0.95-4);
    \draw[thick] (6.5-7,0.22+0.73/2-4) -- (6.5-7,0.95-4);
    \draw[thick] (7-7,1.2-4)node[right]{$=$};
    \node[draw, shape=rectangle, minimum width=5em] (U1) at (8.5-7,1.2+0.44-4){$M$};
    \draw[thick] (8-7,0.22+1.2+0.44-4) -- (8-7,0.95+1.2-0.73/2+0.44-4);
    \draw[thick] (9-7,0.22+1.2+0.44-4) -- (9-7,0.95+1.2-0.73/2+0.44-4);
    \draw[thick] (8-7,0.22+0.73/2+0.44-4) -- (8-7,0.95+0.44-4);
    \draw[thick] (9-7,0.22+0.73/2+0.44-4) -- (9-7,0.95+0.44-4);
    \node[draw, shape=rectangle, minimum width=2em] (U1) at (8-7,0.35+0.44-4){$R$};
    \node[draw, shape=rectangle, minimum width=2em] (U1) at (9-7,0.35+0.44-4){$R$};
    \draw[thick] (8-7,0.22+0.73/2-0.85+0.44-4) -- (8-7,0.95-0.85+0.44-4);
    \draw[thick] (9-7,0.22+0.73/2-0.85+0.44-4) -- (9-7,0.95-0.85+0.44-4);
    \end{tikzpicture}
\caption{Schematic diagram of the trapped ion monitored circuit with open boundary condition. Rectangles with $U$ represent random unitary gates defined in Eq.~(\ref{equ:hybridU}) acting on adjacent qubits. Cyan circles denote local projective measurements along the $Z$-basis with a measurement rate $p$. The size of the whole spin chain is $L$. }
\label{fig:mipt}
\end{figure}
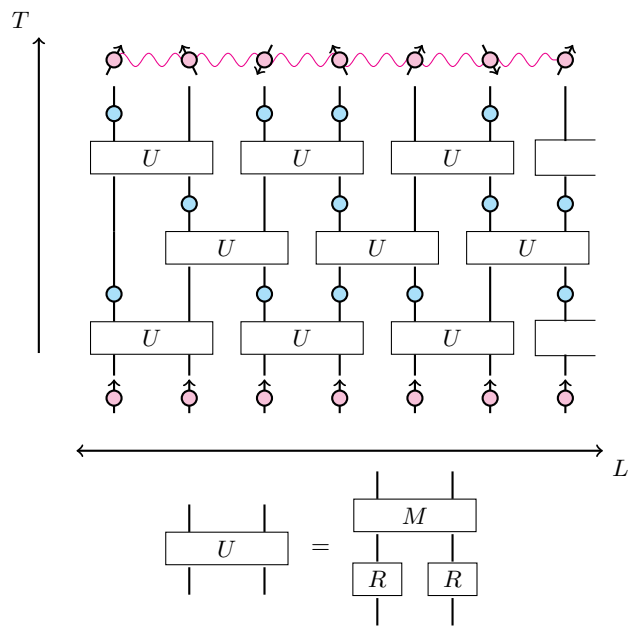

In what follows, the rotation angle in Eq.~(\ref{equ:MSgate}) is fixed as $\theta=\pi/4$. 
For the random single-qubit rotations $R_j$ 
\begin{equation}
    R_j(\theta_j,\varphi_j) ~=~ \begin{pmatrix}
        \cos\frac{\theta_j}{2} & -ie^{-i\varphi_j}\sin\frac{\theta_j}{2} \\
        -ie^{i\varphi_j}\sin\frac{\theta_j}{2} & \cos\frac{\theta_j}{2}
    \end{pmatrix} ~,
    \label{equ:rotation}
\end{equation}
we set $\theta_j=\pi/2$ and each $\varphi_j$ is randomly chosen from one of the following three angles
\begin{equation}
    \varphi_j ~\in~ \left\{ 0~,~ \frac{\pi}{4} ~,~ \frac{\pi}{2} \right\} ~~.
\end{equation}
Note that $R({\pi}/{2},\varphi=0)$ and $R({\pi}/{2},\varphi={\pi}/{2})$ correspond to 90-degree rotations about the $x$- and $y$-axes, respectively. In contrast, $R({\pi}/{2},\varphi={\pi}/{4})$ is not a Clifford gate. 
The random unitaries acting on two adjacent qubits can be expressed collectively as follows
\begin{equation}
        U_{i,i+1}=   
        M_{i,i+1}\!\left(\frac{\pi}{4}\right) R_i\!\left(\frac{\pi}{2},\varphi_i\right) R_{i+1}\!\left(\frac{\pi}{2},\varphi_{i+1}\right)~. 
    \label{equ:hybridU}
\end{equation}
Finally, the complete monitored circuit under study is constructed in a brick-layer fashion as shown in Fig.~\ref{fig:mipt}.  
After each layer of unitaries, we perform single-qubit projective measurements along the $Z$-basis with probability $p$ for each qubit. 

As examined in~\cite{Czischek:2021nso}, random trapped ion hybrid circuits with a circuit depth of $T=4L$ yield steady states, characterized by the convergence of the half-chain entanglement entropy at $T=4L$. Furthermore, by analyzing the dynamical behavior of entanglement entropy across varying measurement rates $p$, \cite{Czischek:2021nso} identified a measurement-induced phase transition in this system. The critical measurement rate was determined to be $p_c\approx 0.17\pm 0.02$.

To obtain stable numerical results for the cross entropy benchmark in the next section, we introduce additional encoding layers following the procedure outlined in~\cite{Li:2022vks}; the circuit with local measurements (Fig.~\ref{fig:mipt}) will be referred to as the bulk. 
Starting with a pure initial product state, we apply encoding layers consisting solely of random unitaries defined in Eq.~(\ref{equ:hybridU}) in the same manner as depicted in Fig.~\ref{fig:mipt}. Subsequently, the bulk circuit is applied. 
With additional encoding layers, the evolution of the half-chain entanglement entropy as a function of time (or circuit depth) is presented in Appendix~\ref{appen:steady}. 
In what follows, we set the encoding and bulk circuit depths to $T_{\rm encoding}=T_{\rm bulk}=2L$. 

Note that this family of random circuits is fully specified by the arrangement of measurements and the choice of unitary gates. 
Thus, in what follows, we represent a circuit $C$ using its corresponding set of $\varphi_i$'s and the positions of measurements $s_i$'s, as follows:
\begin{equation}
    C ~=~ \left\{ \left\{ \varphi^{(t)}_i\right\}_{i=1,\dots,L}^{t=1,\dots,T} ~;~ s_1,\,s_2,\,,\dots,\,s_N \right\}~,
    \label{equ:circuit-params}
\end{equation}
where $T=T_{\rm encoding}+T_{\rm bulk}$ and $s_i=(q_i,t_i)$ represents the measurement site, i.e. specifies the qubit $q_i$ measured at time step $t_i$. Here, $N$ denotes the total number of measurements in the circuit, which varies across different realizations. The sequence $\{s_i=(q_i,t_i)\}$ is ordered by iterating through the spin chain at each time step, repeating the process for subsequent time steps.

\section{Cross entropy benchmark}\label{sec:XEB}

The idea of the cross entropy benchmark for studying the measurement-induced phase transition~\cite{Li:2022vks} is based on the fact that in the volume law phase, the ``coding” property of the system ensures bulk measurements extract little information about the initial state. This is due to strong scrambling in the low-$p$ phase. Conversely, in the area law (high-$p$) phase, some information about the initial state propagates into the measurement outcomes, and to some extent one is able to distinguish different initial states from bulk measurements. Thus the MIPT can be probed by comparing the probability distributions formed by measurement outcomes in circuits with different initial states as follows. 

First, let us set up the notation: For the circuit with an initial state $\rho$, its measurement outcomes $\vec{\mathbf{m}}=\{ m_1,m_2, \dots, m_N\}$ follow a probability distribution and we denote it as $p^{\rho}_{\vec{\mathbf{m}}}$. Now we consider a circuit $C$ and input two different initial states, $\rho$ and $\sigma$. 
The difference between their corresponding probability distributions $p^{\rho}_{\vec{\mathbf{m}}}$ and $p^{\sigma}_{\vec{\mathbf{m}}}$ can be quantified using the cross entropy $\chi_C$ defined in Eq.~(\ref{equ:chiC}). 
The circuit averaged cross entropy $\chi=\mathbb{E}_C\chi_C$ serves as an order parameter for the MIPT. 
In the volume law phase ($p<p_c$), $\chi=1$, indicates that based on measurement outcomes one is not able to distinguish different initial states. In contrast, in the area law phase ($p>p_c$), $\chi$ takes on a constant value less than 1, signifying that information about the initial states is accessible from the bulk measurements.

The cross entropy for a given circuit $C$ is defined as follows
\begin{equation}
        \chi_{C} ~=~\frac{\sum_{\vec{\mathbf{m}}}\,p^{\rho}_{\vec{\mathbf{m}}}\,p^{\sigma}_{\vec{\mathbf{m}}}}{\sum_{\vec{\mathbf{m}}}\,p^{\sigma}_{\vec{\mathbf{m}}}\,p^{\sigma}_{\vec{\mathbf{m}}}} ~=~ \frac{\Big\langle p^{\sigma}_{\vec{\mathbf{m}}}\Big\rangle_{\rho}}{\Big\langle p^{\sigma}_{\vec{\mathbf{m}}}\Big\rangle_{\sigma}} ~~, 
        \label{equ:chiC}
    \end{equation}
where in the first equality, the sums run over all possible measurement outcome $\vec{\mathbf{m}}$'s. For a generic circuit, the total number of possible outcomes grows exponentially with the system size and measurement rate, scaling as $2^{pLT}$. 
Due to this exponential growth, directly evaluating the cross entropy is computationally infeasible. Instead, we approximate the numerator and denominator independently using the law of large numbers~\cite{Li:2022vks}.
Namely,
\begin{equation}
    \begin{split}
        \Big\langle p^{\sigma}_{\vec{\mathbf{m}}}\Big\rangle_{\rho} ~=&~  \lim_{M\to\infty}\,\frac{1}{M}\,\sum_{j=1, \vec{\mathbf{m}}_j \sim p^{\rho}_{\vec{\mathbf{m}}}}^M\,p^{\sigma}_{\vec{\mathbf{m}}_j}  ~, \\
        \Big\langle p^{\sigma}_{\vec{\mathbf{m}}}\Big\rangle_{\sigma} ~=&~ \lim_{M\to\infty}\,\frac{1}{M}\,\sum_{k=1, \vec{\mathbf{m}}_k \sim p^{\s}_{\vec{\mathbf{m}}}}^{M}\,p^{\sigma}_{\vec{\mathbf{m}}_k}  ~,
    \end{split}
    \label{equ:estimate_chi}
\end{equation}
where $\vec{\mathbf{m}}_j \sim p^{\rho}_{\vec{\mathbf{m}}}$ means that the outcome $\vec{\mathbf{m}}_j$ is sampled from the probability distribution $p^{\rho}_{\vec{\mathbf{m}}}$.

With the help of classical simulations, \cite{Li:2022vks} proposed an experimental protocol for observing MIPT on a quantum device. The procedure is outlined as follows. 
(1) Run the $\rho$-circuit: Initialize a quantum computer with the state $\rho$ and execute the random circuit. Record the circuit parameters $C$ defined in Eq.~(\ref{equ:circuit-params}) and measurement outcomes $\vec{\mathbf{m}}$. 
(2) Simulate the $\sigma$-circuit: on a classical computer, simulate the circuit $C$ with a different initial state $\sigma$, replacing the projective measurements with their corresponding projector operators applied based on $\vec{\mathbf{m}}$ (without normalization). Thus the probability $p^{\sigma}_{\vec{\mathbf{m}}}$ for obtaining measurement result $\vec{\mathbf{m}}$ in the $\sigma$-circuit is simply the trace of the final state density matrix. For pure states, this corresponds to the norm of the final state. In what follows, we consider\,\footnote{These initial states can be prepared with high fidelities in trapped-ion systems using optical pumping~\cite{RevModPhys.44.169}, with fidelity $>99.9\%$~\cite{Christensen:2020dik}, and single-qubit rotation gates with fidelity $>99.9999\%$~\cite{Smith:2024gbs} applied globally.}
\begin{equation}
    \rho ~=~|+\rangle^{\otimes L}~,~ 
    \sigma ~=~|0\rangle^{\otimes L}~.
\end{equation}
(3) Compute $\chi_C$: perform $M$ measurement runs on the quantum computer using the same circuit $C$. Compute the numerator in Eq.~(\ref{equ:chiC}) as the mean of $p^{\sigma}_{\vec{\mathbf{m}}}$ across all runs. Similarly, calculate the denominator in Eq.~(\ref{equ:chiC}) and use these results to determine the cross entropy $\chi_C$ for the circuit $C$. 
(4) Average over circuits: Repeat the procedure for $M_C$ sampled $\rho$-circuits and compute the average of $\chi_C$ across these circuits.

This protocol is scalable for stabilizer circuits~\cite{Li:2022vks}, while for generic non-stabilizer circuits, the scalability is restricted by classical simulations. Additionally, the sample complexity\textemdash namely, the number of measurement runs $M$ required to obtain an accurate estimation of the cross entropy as defined in~\cite{Iouchtchenko:2022php}\textemdash is relatively unexplored. In this work, we take the first step toward addressing it. 
As shown in Eq.~(\ref{equ:chiC}) and (\ref{equ:estimate_chi}), the XEB relies solely on measurement outcomes and their associated probabilities. This feature naturally aligns with the framework of language models, as we will see in the next section. 
Before delving into the implementation and advantages of using language models, in the rest of this section, we validate the XEB for the hybrid circuit of interest introduced in Sec.~\ref{sec:circuit} using classical simulations.

\begin{figure}[t]
    \centering
    \includegraphics[width=\linewidth]{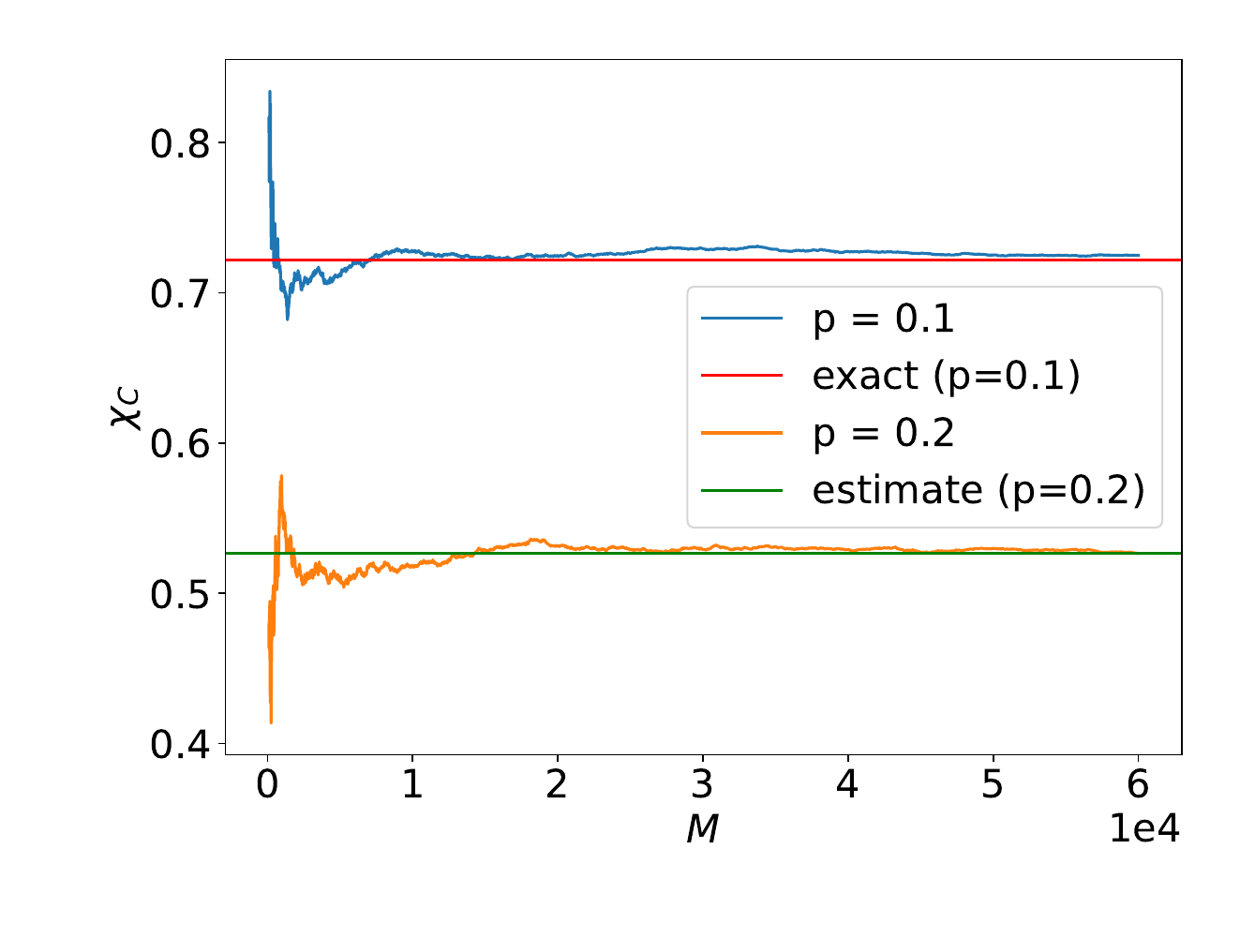}
    \caption{For two given circuits with $L=8$ and different $p$'s, the evolution of the cross entropy $\chi_C$ estimated using Eq.~(\ref{equ:estimate_chi}) as a function of the number of measurement runs $M$. 
    For $p=0.1$, the exact value of $\chi_C$ is plotted in red. For $p=0.2$, the estimation of $\chi_C$ at $M=6\times 10^5$ is plotted in green for reference.}
    \label{fig:benchmark_M}
\end{figure}

Note that in stabilizer circuits, when both initial states are stabilizer states, $\chi_C$ can be computed exactly since Eq.~(\ref{equ:estimate_chi}) has a closed form. Effectively, this corresponds to having an infinite number of measurement runs, $M=\infty$. However, for the trapped ion hybrid circuits, which are non-stabilizer circuits, it is necessary to first study the behavior of the cross entropy $\chi_C$ for a given circuit as a function of the number of measurement runs $M$. To do so, we consider a small system size $L=8$ and simulate two circuits generated with measurement rates $p=0.1$ and $p=0.2$ respectively. The corresponding number of projective measurements is $N=12$ for $p=0.1$ and $N=24$ for $p=0.2$. 
For the $p=0.1$ circuit, the relatively small $N$ allows us to compute the exact value of $\chi_C$ by summing over all $2^N=4096$ possible measurement outcomes as in Eq.~(\ref{equ:chiC}). This exact value will serve as a benchmark to evaluate the accuracy of the estimations in the next section. 
For the $p=0.2$ circuit, we focus solely on the trend of $\chi_C$. As shown in Fig.~\ref{fig:benchmark_M}, in both cases, $\chi_C$ initially fluctuates before gradually converging to a constant value as $M$ increases.  
In the following simulations, we set $M=5\times 10^3$ to compute the cross entropy $\chi_C$ for each circuit using Eq.~(\ref{equ:estimate_chi}). This value is selected based on convergence trends: despite being relatively small (to save computational resources), it provides a reliable estimation.\,\footnote{We have also verified the $\chi_C$-$M$ dependence for $L=10$ and $L=16$ and various values of $p$. In all cases, $M=5\times 10^3$ is an appropriate choice. }  

\begin{figure}[t]
    \centering
    \includegraphics[width=1.0\linewidth]{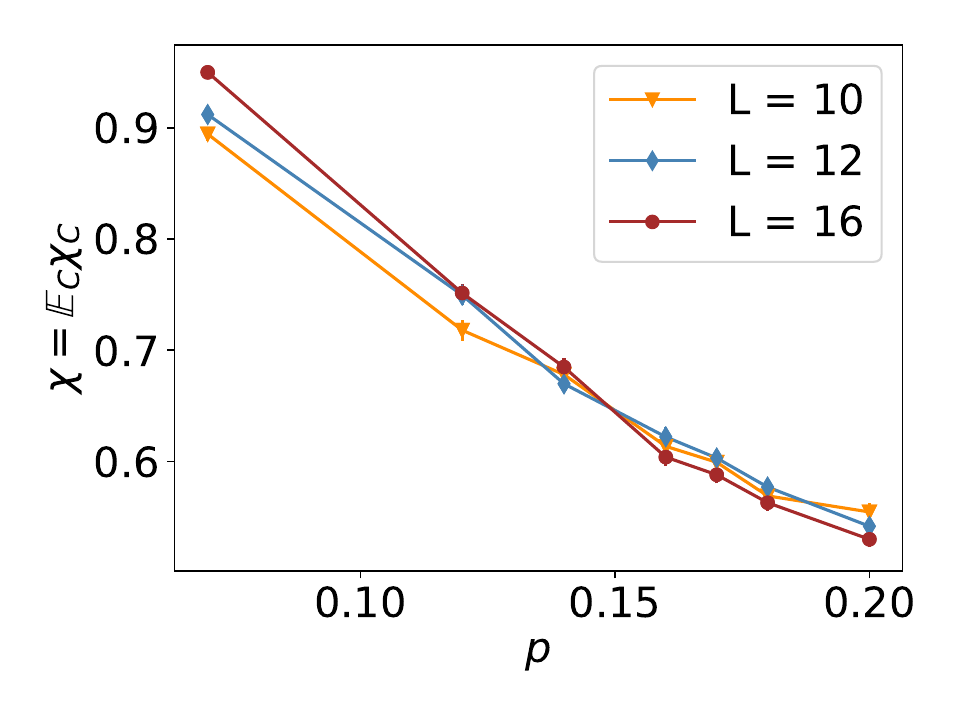}
    \caption{Numerical results of the circuit-averaged cross entropy $\chi$ for trapped ion hybrid circuits with initial states $\rho=|+\rangle^{\otimes L}$ and $\sigma=|0\rangle^{\otimes L}$. For each circuit, we estimate the cross entropy $\chi_C$ following Eq.~(\ref{equ:estimate_chi}) and using $M=5\times 10^3$ measurement runs. This value of $M$ is selected based on convergence trends. 
    Then we average over $M_C=100$ circuits to estimate $\chi$. Error bars represent one standard deviation of the cross entropy, averaged over 100 independent circuits. 
    }
    \label{fig:chi_p}
\end{figure}

We next simulate hybrid circuits with various system sizes $L$ and compute the cross entropy $\chi$ following the procedure outlined above. In particular, for each system size, we simulate $M_C=100$ circuits, and for each circuit configuration, we sample $M=5\times 10^3$ measurement outcomes for both $\rho$- and $\sigma$-circuits. 
The numerical results are shown in Fig.~\ref{fig:chi_p}, which perform similar behavior as that of Haar random circuits for small system sizes examined in~\cite{Li:2022vks}. 
Ideally, in the limit where $M$, $M_C$, and $L$ approach infinity, $\chi$ would behave as a step function of the measurement rate $p$: $\chi=1$ for $p<p_c$ and dropping to a constant at the critical point $p_c$. However, due to finite-size effects, the $p-\chi$ curves for different $L$'s are expected to exhibit a downward slope while intersecting at the critical point, which is indeed observed in Fig.~\ref{fig:chi_p}. 
This crossing behavior confirms that the XEB is effective for trapped ion hybrid circuits and the critical measurement rate is estimated to be $p_c\approx 0.15$ based on the plot, which is consistent with~\cite{Czischek:2021nso}. 

\section{Recurrent Neural Network Implementation}\label{sec:RNN}

One core aspect of language models is their ability to process sequences of text or data in a probabilistic manner. Given the preceding context, these models generate the next token in a sequence based on a learned conditional probability distribution. Modern architectures for sequence processing, such as recurrent neural networks (RNNs) and transformers, have been studied extensively on their applicability to representing quantum states~\cite{Carrasquilla_2019,hibat2020recurrent,hou2023quantum,Melko2024-ev,Fitzek:2024onn,Yang:2024yxu}. In this work, we apply RNNs to learn the probability distributions for both $\rho$- and $\sigma$-circuits using the measurement outcomes generated via classical simulations in Sec.~\ref{sec:XEB}. Specifically, we focus on the same two circuits with $L=8$ simulated to obtain results in Fig.~\ref{fig:benchmark_M}. In what follows, the estimation method described below Eq.~(\ref{equ:estimate_chi}) is referred to as the histogram approach. 

Below we first introduce the RNN model for learning probability distributions in Sec.~\ref{subsec:RNN} and then demonstrate its implementation in the cross entropy benchmark in Sec.~\ref{subsec:implementation}. Finally, in Sec.~\ref{subsec:training}, we analyze the data complexity by comparing the number of measurement runs required for accurate estimation of $\chi_C$ using the histogram approach versus RNNs.

\subsection{Vanilla RNNs for probability distributions}\label{subsec:RNN}

\begin{figure*}[t]
    \centering
    \subfloat[\label{fig:rnn-a}]{%
    \includegraphics[width=0.47\textwidth]{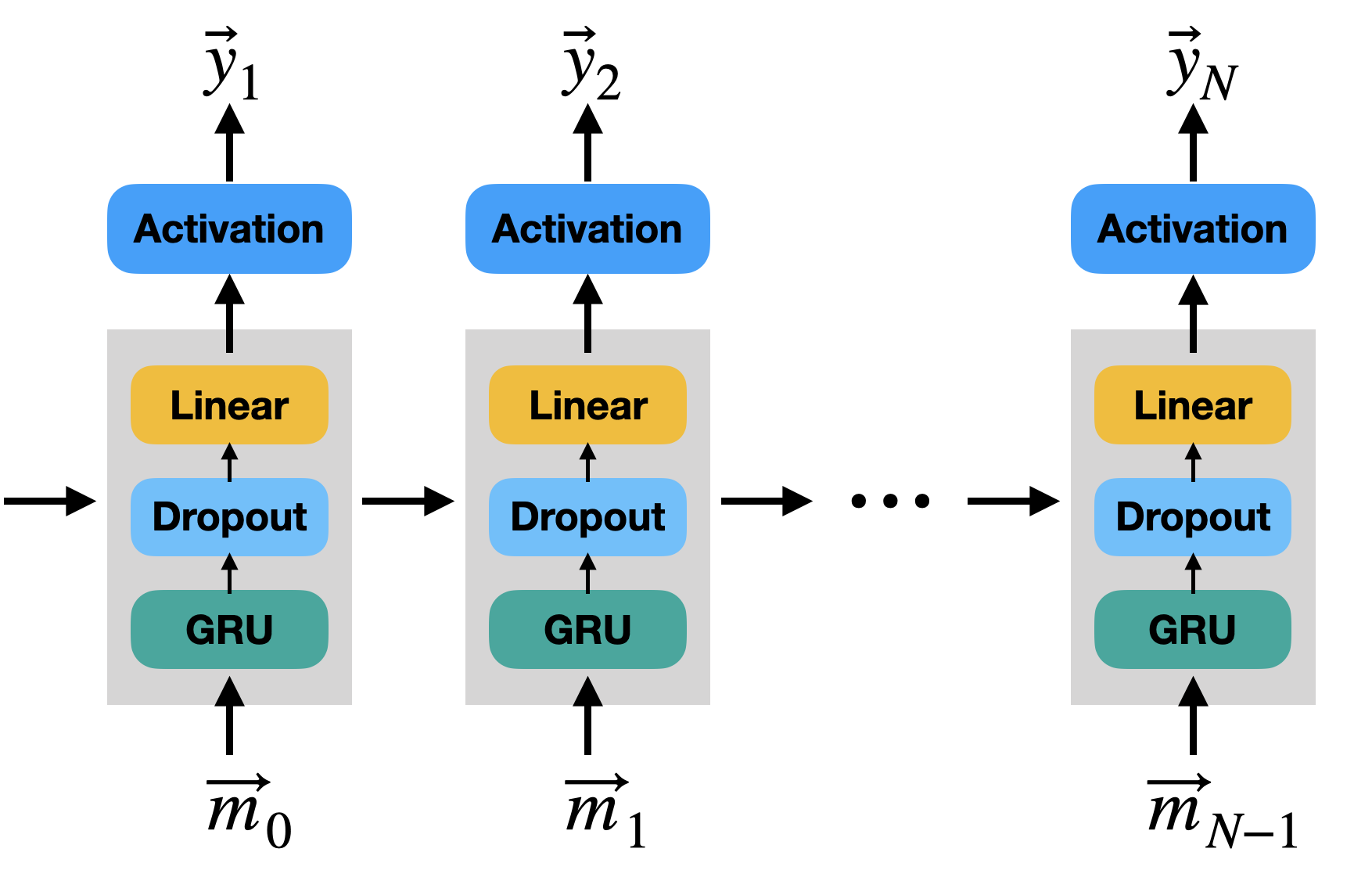}}
    \hfill
    \subfloat[\label{fig:rnn-b}]{%
    \includegraphics[width=0.45\textwidth]{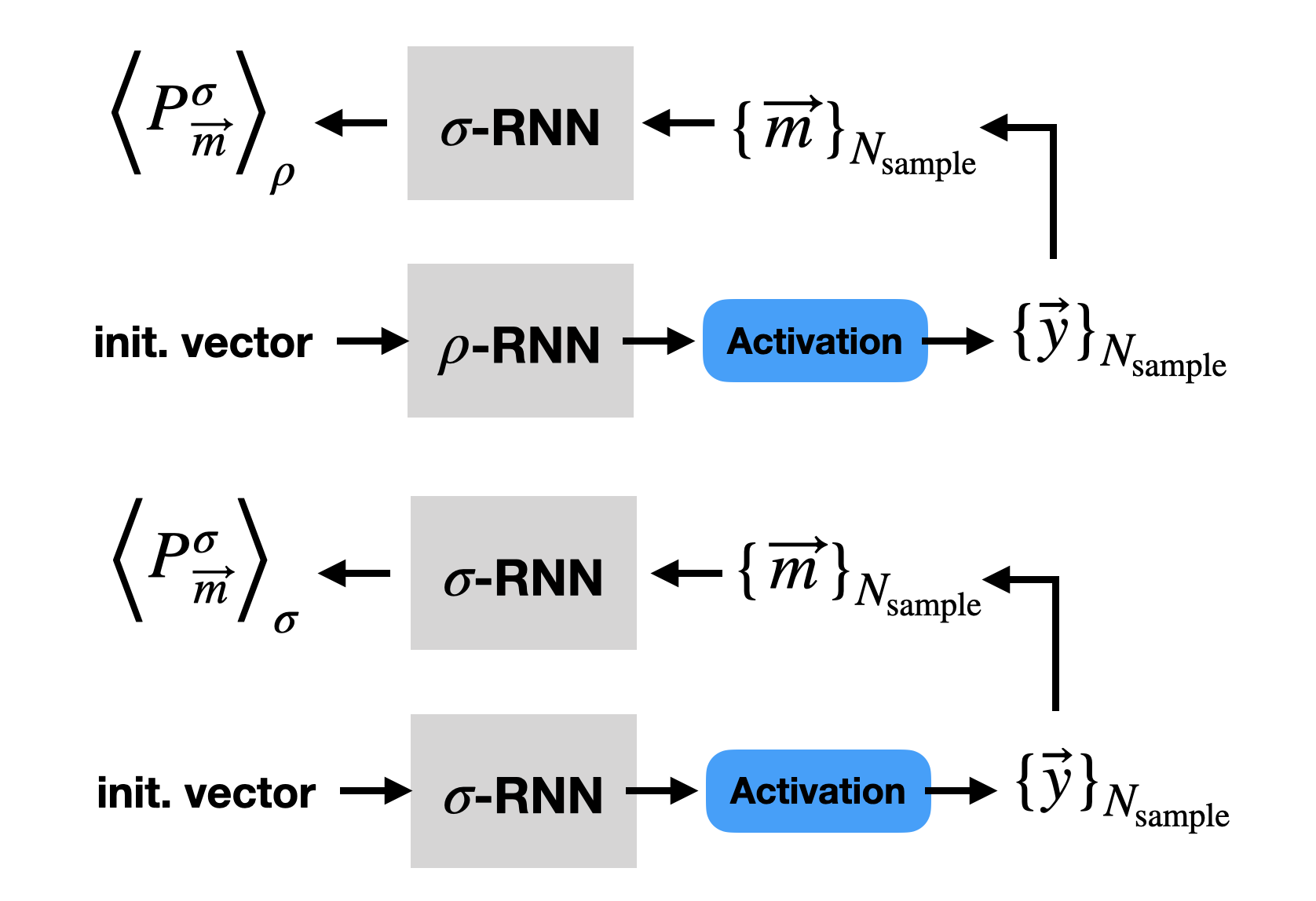}}
    \caption{(a): A schematic diagram of the RNN model used in this work. The gray-shaded block represents the RNN cell, which processes a sequence of inputs $\{\vbm\}$, where each input vector has a dimension of $N_i=2$. Here $\vec{m}_0=(0,0)$ is an initial vector to kick off the model. 
    Within the RNN cell, the GRU layer generates a hidden state with a dimension of $N_h$. After passing through a Softmax activation layer, the output vector $\vec{y}_i$ (dimension $N_o=2$) corresponds to the conditional probability distribution, as described in Eq.~(\ref{equ:cond_prob}). 
    (b): The RNN estimation procedure for the numerator and denominator in Eq.~(\ref{equ:chiC}). The autoregressive property of the model yields a recursive sampling process as follows. An initial zero vector $\vec{m}_0$ is fed into the RNN, producing the conditional probability $\vec{y}_1$ for the first site. Based on this probability, a configuration is sampled at the first site. This outcome is then used to sample the second site based on $\vec{y}_2$, and so on.}
    \label{fig:rnn}
\end{figure*}

We consider a similar architecture for simple (vanilla) RNNs as in~\cite{hibat2020recurrent}.\,\footnote{For a comprehensive overview of vanilla RNNs, including details on activation functions and parameters, we refer readers to~\cite{hibat2020recurrent}.} The schematic diagram of this model is presented in Fig.~\ref{fig:rnn-a}. 

The input data is a one-hot encoding of the measurement outcomes $\vbm=\{m_1,m_2,\dots,m_{N}\}$ in the monitored circuit. Here, $N$ is the total number of measurements in a given circuit $C$, as defined in Eq.~(\ref{equ:circuit-params}), and each measurement outcome $m_i$ takes a value of either 0 or 1 for qubits. Through one-hot encoding, $0$ and $1$ are mapped to $(1,0),(0,1)$ respectively, making the input dimension $N_i=2$. We use the vector notation $\vec{m}_i$ to denote the one-hot encoding of $m_i$. 
As shown by the gray-shaded block in Fig.~\ref{fig:rnn-a}, the building block of this model, called the RNN cell, comprises the following: (1) a gated recurrent unit (GRU) layer\,\footnote{A GRU~\cite{cho-etal-2014-properties} layer employs gating mechanisms, including the reset and update gates, to selectively retain and update information from preceding inputs. 
It is the key component that enables the autoregressive feature in this RNN.} that generates a hidden state with dimension $N_h$;
(2) a dropout layer for regularization to prevent overfitting; (3) a linear layer that maps hidden states to output logits with a dimension of $N_o=2$. 
Subsequently, a Softmax activation layer converts the logit into $\vec{y}_i$, corresponding to the conditional probability distribution of the outcomes at site $i$. Namely, $\vec{y}_i$ is a two-dimensional vector with non-negative real components that sum to one. Given the preceding measurement record $\{m_1,\dots,m_{i-1}\}$, the conditional probability of obtaining $m_i$ at site $i$ reads
\begin{equation}
    P(m_i|m_1,m_2,\dots,m_{i-1}) ~=~ \vec{y}_i\,\cdot\,\vec{m}_i  ~,
    \label{equ:cond_prob}
\end{equation}
where $\cdot$ is the inner product of two 2-dimensional vectors. 
The probability of observing the full configuration $\vbm$ is the production of a sequence of these conditional probabilities. Namely,  
\begin{equation}
    P(m_1,m_2,\dots,m_{N}) ~=~ \prod_{i=1}^{N}\,\vec{y}_i\,\cdot\,\vec{m}_i ~.
    \label{equ:P(m)}
\end{equation}

To learn the probability distribution $p_{\vbm}$, we train the model using a set of outcome configurations $\{\vbm\}$ sampled from $p_{\vbm}$. We employ the Negative Log Likelihood as the loss function
\begin{equation}
    {\cal L} ~=~ - \frac{1}{N}\, {\rm avg}_{\{\vbm\}}\, \log P(\vbm) ~,
\end{equation}
where the loss is averaged over both the training batches and the measurement sites. This loss function evaluates how closely the model’s predicted probability distribution matches the true one.

\subsection{RNN-enhanced protocol}\label{subsec:implementation}

The model introduced above is capable of learning any classical probability distribution. In the context of the cross entropy benchmark, RNNs can play the role of a decoder, mapping the measurement outcomes in a monitored circuit $C$ initialized with two different states to its corresponding cross entropy $\chi_C$. The RNN-enhanced experimental protocol is outlined as follows. 
\begin{enumerate}
    \item Run $\rho$- and $\sigma$-circuits: Sample a random circuit $C$ with a given system size $L$ and measurement rate $p$. Execute this circuit $C$ independently with initial states $\rho$ and $\sigma$ on quantum computers, recording the measurement outcomes for both $\rho$- and $\sigma$-circuits.
    \item Train $\rho$- and $\sigma$-RNNs: Use the respective sets of measurement outcomes to train two separate vanilla RNNs. The first, denoted as $\rho$-RNN, learns the probability distribution $p^{\rho}_{\vbm}$, while the second, called $\sigma$-RNN, learns $p^{\sigma}_{\vbm}$.
    \item Compute $\chi_C$: With the trained RNNs, sample $N_{\rm sample}$ outcomes $\{\vec{m}\}_{N_{\rm sample}}$ from $\rho$-RNN and feed them into $\sigma$-RNN. The mean of the output probabilities, defined in Eq.~(\ref{equ:P(m)}), provides an estimate of the numerator in Eq.~(\ref{equ:chiC}). Similarly, calculate the denominator in Eq.~(\ref{equ:chiC}) and obtain $\chi_C$. This estimation procedure is depicted in Fig.~\ref{fig:rnn-b}. 
    \item Average the cross entropy over circuits. 
\end{enumerate}

This protocol replaces the classical simulations in~\cite{Li:2022vks} with trained RNNs. Consequently, the scalability of this approach hinges on the scalability of the RNNs, which will be discussed in Sec.~\ref{sec:discussion}. In the following, we demonstrate the advantages of this protocol by analyzing the data complexity.

\subsection{Training results and data complexity}\label{subsec:training}

\begin{figure}[htp!]
    \centering
    \subfloat[\label{fig:compare-a}]{%
    \includegraphics[width=0.33\textwidth]{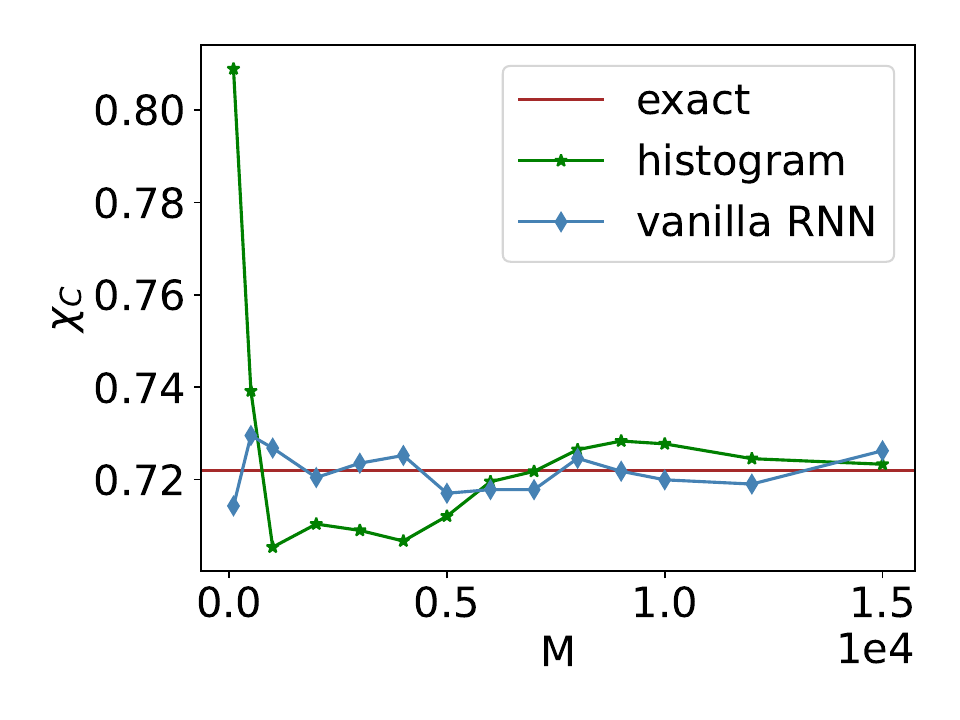}}
    \subfloat[\label{fig:compare-b}]{%
    \includegraphics[width=0.16\textwidth]{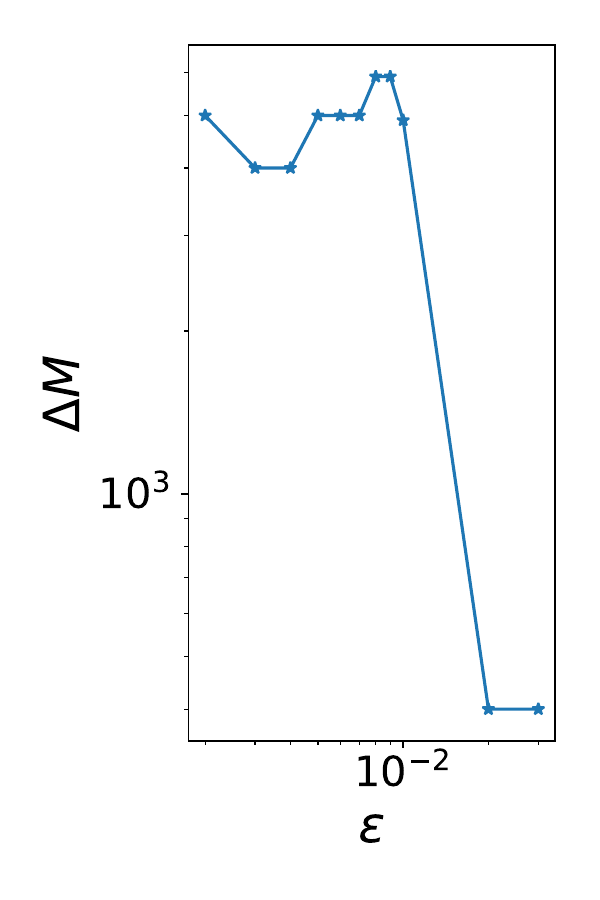}}
    \caption{The performance of the $\chi_C$ estimation for the circuit with $L=8$ and $p=0.1$ using the histogram approach versus vanilla RNNs. (a): the evolution of $\chi_C$ as a function of $M$. The exact result of $\chi_C$ is plotted in red. (b): $\Delta M$ defined in Eq.~(\ref{equ:DeltaM}) as a function of the accuracy $\varepsilon$ defined in Eq.~(\ref{equ:accuracy}). \label{fig:compare}}
\end{figure}

\begin{figure}[htp!]
    \centering
    \subfloat[\label{fig:compare-p02-a}]{%
    \includegraphics[width=0.33\textwidth]{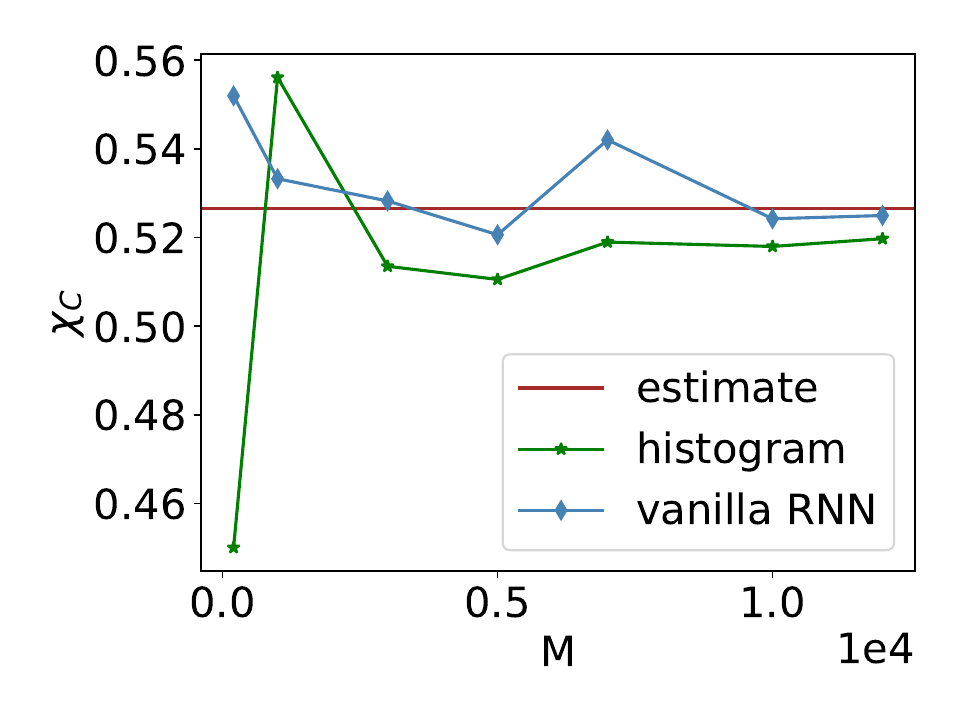}}
    \subfloat[\label{fig:compare-p02-b}]{%
    \includegraphics[width=0.16\textwidth]{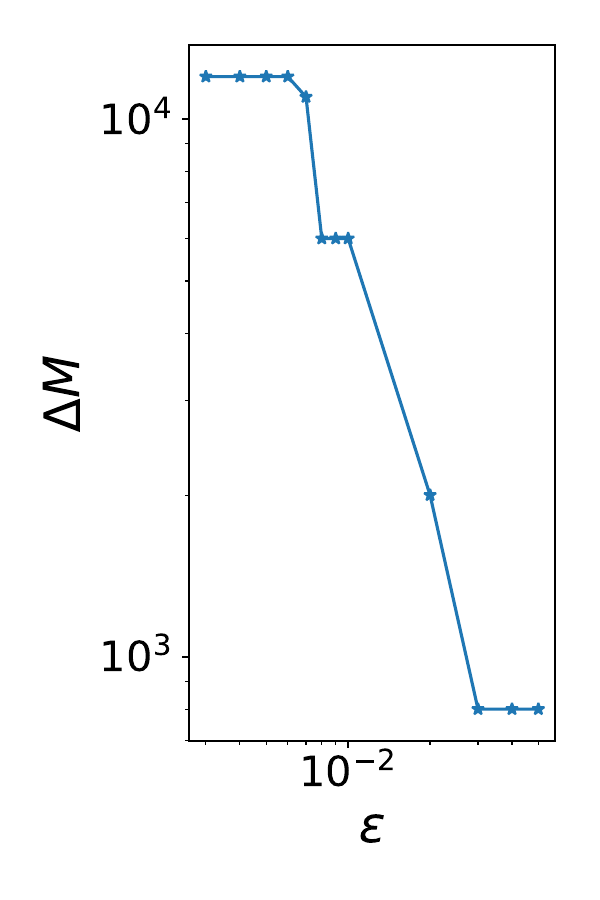}}
    \caption{The performance of the $\chi_C$ estimation for the circuit with $L=8$ and $p=0.2$ using the histogram approach versus vanilla RNNs. (a): the evolution of $\chi_C$ as a function of $M$. The histogram result at $M=6\times 10^5$ is plotted in red for reference. (b): $\Delta M$ defined in Eq.~(\ref{equ:DeltaM}) as a function of the accuracy $\varepsilon$. \label{fig:compare-p02}}
\end{figure}

As guiding examples, we revisit the two circuits with system size $L=8$ and measurement rates $p=0.1$ and $p=0.2$ studied in Sec.~\ref{sec:XEB}. To analyze the data complexity, we compare the evolution of estimated $\chi_C$ as a function of the number of measurement runs $M$ using the histogram approach versus RNNs. For each value of $M$, the same set of measurement outcomes, generated via classical simulations, is used to train the RNNs. 
The results for $p=0.1$ and $p=0.2$ are shown in Fig.~\ref{fig:compare-a} and Fig.~\ref{fig:compare-p02-a} respectively.\,\footnote{Although we focus on two example circuits here, similar results can be obtained for other circuit realizations.} 
The corresponding RNN model parameters are listed in Appendix~\ref{appen:RNN}. 
In both cases, the values of $\chi_C$ obtained using the RNN and histogram approaches fluctuate around a constant value at which they converge. The RNN-enhanced protocol exhibits smaller fluctuations, highlighting its advantages, which we will quantify shortly. 

Recall that the total number of measurements $N$ is $N=12$ for $p=0.1$ and $N=24$ for $p=0.2$. For the $p=0.1$ circuit, we have computed the exact value of $\chi_C$, enabling us to define the accuracy $\varepsilon$ of the estimation as the difference between the exact and estimated values:
\begin{equation}
    \varepsilon ~:=~ |\chi_C^{\rm exact} - \chi_C| ~.
    \label{equ:accuracy}
\end{equation}
To assess the performance of each estimation approach, we determine the minimum number of measurement runs $M_{\rm min}$ required to achieve a target accuracy $\varepsilon$.\,\footnote{This is an analog of the sample complexity defined in~\cite{Iouchtchenko:2022php}.} The improvement offered by the RNN-enhanced protocol is quantified by the relative reduction in measurement runs:
\begin{equation}
    \Delta M ~:=~ M_{\rm min}^{\rm histogram} ~-~ M_{\rm min}^{\rm RNN} ~,
    \label{equ:DeltaM}
\end{equation}
which is analyzed for various accuracy targets $\varepsilon$ in Fig.~\ref{fig:compare-b}. 
For the $p=0.2$ circuit, where calculating the exact value of $\chi_C$ is infeasible, we treat the steady value of $\chi_C$ at $M=6\times 10^5$ as the exact value for the purpose of evaluating accuracy. 
The result is shown in Fig.~\ref{fig:compare-p02-b}. 
In both cases (Fig.~\ref{fig:compare-b} and Fig.~\ref{fig:compare-p02-b}), $\Delta M$ is always positive, indicating that the RNN-enhanced protocol outperforms the histogram approach. 
In particular, it significantly reduces the number of measurement runs required to achieve a given accuracy. For high accuracy (i.e. small $\varepsilon$), the improvement is substantial, with reductions on the order of $10^3$ for $p=0.1$ and $10^4$ for $p=0.2$. As the accuracy requirement decreases (i.e. $\varepsilon$ increases), $\Delta M$ also drops. This aligns with intuitions: When accuracy is relaxed, both $M_{\rm min}^{\rm histogram}$ and $M_{\rm min}^{\rm RNN}$ are reduced, resulting in a smaller difference between them.
Namely, the stricter the accuracy requirement, the greater the advantage of the RNN-enhanced protocol over the histogram approach.

\section{Discussion}\label{sec:discussion}

In this work, we examine the cross entropy benchmark (XEB) for a measurement-induced phase transition in trapped ion monitored circuits. We propose a recurrent neural network (RNN) strategy to enhance the XEB signal obtained in a data-limited setting, with the goal of exploring how generative machine learning models can assist in quantum computer benchmarking and control. We'll close with several remarks and natural routes for future investigations. 

When evaluating the computational efficiency of a machine learning protocol, two primary aspects are typically considered: computational resources and sample complexity. In this work, our focus has been on the latter, which is crucial in the case of measurement-induced phase transitions, since the calculation of the XEB requires knowledge of the probability distribution underlying the measurement record. We have demonstrated that using an RNN-based generative model can improve the data efficiency of the XEB calculation, as compared to using a parameter-free distribution simply obtained from the data frequency counts. In this paper, we have focussed on a demonstration using vanilla RNNs on two specific circuits with simplified parameter settings. 
A more systematic assessment of the RNN-enhanced protocol in terms of sample complexity is left for future work. 

Like in many machine learning strategies, the power of generative models lies in their broad applicability, scalability, and ability to generalize well to unseen data.
The scalability of our RNN-enhanced protocol depends on extending the generative models to handle a larger number of qubits $N$. 
Among generative models, autoregressive language models like the RNN are particularly promising in this regard due to their demonstrated abilities to scale. 
Notably, the current state-of-the-art language model is the attention-based transformer~\cite{vaswani2017attention,PhysRevB.107.075147,lange2024transformer,Fitzek:2024onn,Zhang_2024,yao2024shadowgpt,Wang:2024pdt}. Further investigation is necessary to evaluate different autoregressive models and assess their efficiency in the context of our problem.

Regarding potential generalizations in the model architecture, one natural direction would be to incorporate the circuit parameters as an encoder, enabling the machine learning model to learn an entire family of circuits rather than being restricted to a single probability distribution. 
For instance, the current input data is a flattened sequence of measurement outcomes without any information about the measurement positions. 
To better capture spatial and temporal correlations, as in~\cite{Dehghani:2022aia}, one could adopt the input as an $L \times T$ matrix that records the measurement configuration at each site: entries of $0$ indicate no measurement, while $\pm1$ correspond to the measurement outcomes.

Note that in this work we use synthetic data to train the RNN models. As a next step toward benchmarking quantum devices, it would be interesting to introduce noise models in the simulations. One natural starting point could be incorporating a natural noise model for the trapped ion devices discussed in~\cite{Czischek:2021nso}. Additionally, conducting simulations with larger system sizes and training the models on real experimental data will be key to demonstrating the effectiveness of our protocol in the future.

\section*{Acknowledgements}
Numerical simulations and machine learning in this work were enabled in part by support provided by the Digital Research Alliance of Canada (\url{alliancecan.ca}).
The learning results were produced using the Pytorch package~\cite{paszke2019pytorch}. 
YH is supported by an Alliance Quantum Program grant funded by NSERC, for the project entitled ``A new generation of hardware efficient superconducting qubits". YH thanks Zi-Wen Liu for the valuable discussions. 
WWK is supported by
a grant from the Fondation Courtois, a Chair of the Institut Courtois, a Discovery Grant from NSERC, and a Canada Research Chair.
RGM is supported by NSERC and the Perimeter Institute for Theoretical Physics.
Research at the Perimeter Institute is supported by the Government of Canada through the Department of Innovation, Science and Industry Canada and by the Province of Ontario through the Ministry of Colleges and Universities.

\appendix 

\section{Benchmark for the circuit depth}\label{appen:steady}
In this appendix, we study the evolution of the half-chain von Neumann entropy to determine the time (or equivalently the circuit depths) required for reaching steady states in the encoding-bulk circuits described in Section~\ref{sec:circuit}. 

\begin{figure}[tbp]
    \centering
    \includegraphics[width=0.5\textwidth]{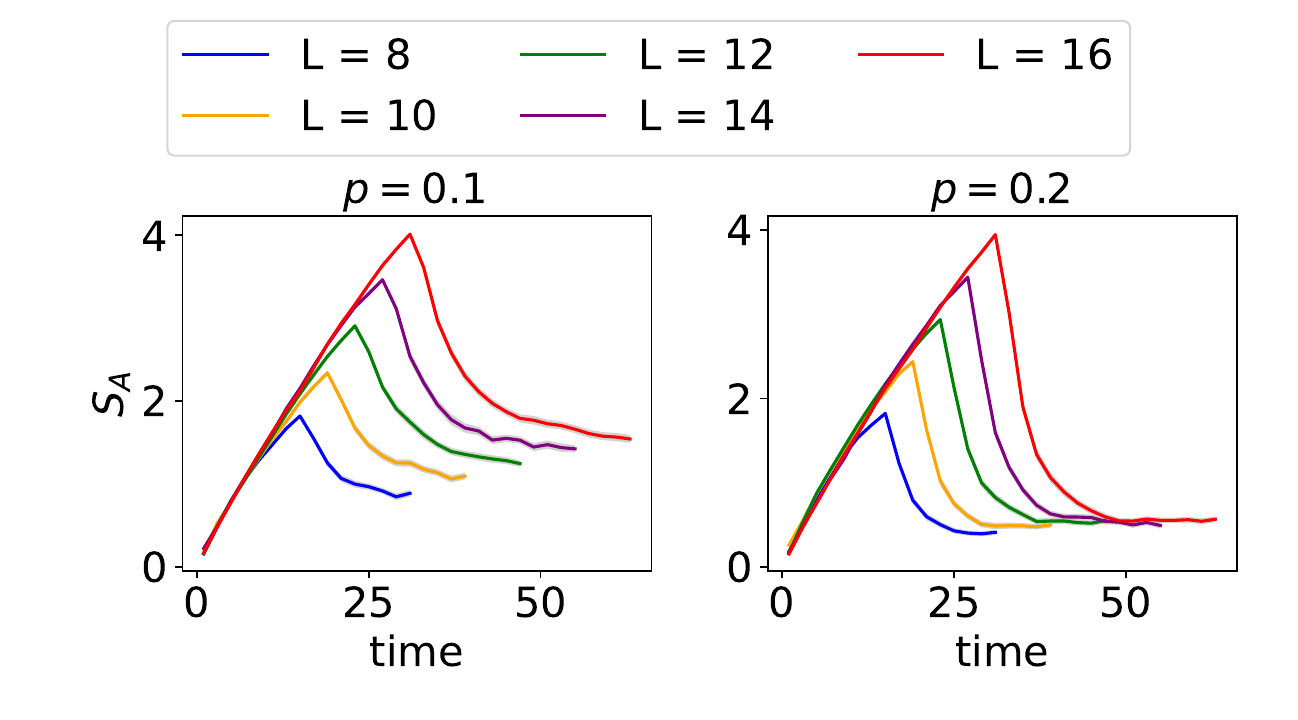}
    \caption{The half-chain von Neumann entropy as a function of time for various system sizes $L$ and measurement rates $p = 0.1$ and $p = 0.2$. For each curve in the plots, 100 circuits are averaged and the standard mean error is shown in the gray-shaded area.}
\label{fig:SvN_evolution_encoding}
\end{figure}

The von Neumann entropy for a subsystem $A$ is defined as follows
\begin{equation}
    S_A ~=~ -{\rm Tr}(\rho_A\log \rho_A) ~,
\end{equation}
where $\rho_A={\rm Tr}_{\bar{A}}\rho$ is the reduced density matrix obtained by tracing out the complement of the subsystem $A$. For this study, $A$ is chosen as the first $L/2$ qubits. Since we focus on small system sizes in this work, $S_A$ is simply computed using the exact diagonalization of the density matrix. 
The time evolution of the half-chain entanglement entropy for $L=8, 10, 12, 14, 16$ is shown in Figure~\ref{fig:SvN_evolution_encoding}. 
Here we consider two measurement rates: $p=0.1$ and $p=0.2$. For each case, we simulate $M_C = 100$ circuits, and their half-chain entanglement entropies $S_A$ are averaged over these independent circuits. 
Moreover, note that the circuits follow a brick-wall layout, distinguishing between even and odd time steps. To further smooth the results, we average $S_A$ over adjacent odd and even time steps. 

As shown in Fig.~\ref{fig:SvN_evolution_encoding}, for all systems considered, the half-chain von Neumann entropy $S_A$ first increases, then decreases, and eventually converges to a constant value. This behavior is consistent with the encoding-bulk circuit architecture as follows. In the encoding period, entangling unitaries continuously scramble the information, causing $S_A$ to increase. After turning on the measurements in the bulk period, occasional measurements lower the $S_A$. 
Moreover, Fig.~\ref{fig:SvN_evolution_encoding} demonstrates that a circuit depth of $T_{\rm encoding}=T_{\rm bulk}=2L$ is sufficient to generate steady states; therefore, we adopt $T_{\rm encoding}=T_{\rm bulk}=2L$ in the simulations presented in Sec.~\ref{sec:XEB}. 

\section{RNN models}\label{appen:RNN}

This appendix provides the model parameters for the RNNs used in Sec.~\ref{subsec:training}. For the $p=0.1$ circuit, the parameters are listed in Table~\ref{tab:vrnn_params}, and those for the $p=0.2$ circuit are listed in Table~\ref{tab:vrnn_params-p02}. 
For simplicity, at each $M$, the same set of model parameters is used for both $\rho$- and $\sigma$-RNNs, with Pytorch's default initialization applied~\cite{paszke2019pytorch}. 
The parameters listed in the tables are explained as follows.

In both cases, the learning rate is fixed at $10^{-3}$ throughout the training process. 
In the training, the dataset with size $M$ is divided into a training set and a validation set. During each training epoch, the dataset is further split into batches to calculate the gradient of batch loss with respect to all parameters. The batch size and the validation set size are listed in the tables. 
As defined in Sec.~\ref{subsec:RNN}, $N_h$ denotes the dimension of the hidden state, and the dropout rate for the dropout layer is also provided in the tables. 
As introduced in Sec.~\ref{subsec:implementation}, $N_{\rm sample}$ is the number of configurations we sample from the trained RNNs to compute the cross entropy. Recall that for the $p=0.1$ circuit, we compute the $\chi_C$ by summing over all possible configurations as in Eq.~(\ref{equ:chiC}). Namely, $N_{\rm sample}=\infty$. 
The training time, represented as the total number of training epochs, is denoted as $N_{\rm epochs}$ in the tables. 
\begin{table}[htp!]
    \centering
    \begin{tabular}{c|c|c|c|c}
       $M$ & batch size & validation size &  $N_h$  & dropout rate  \\\hline
       15000 & 1000 & 3000 & 20 & 0.2  \\
       12000 & 1000 & 2000 & 18 & 0.2  \\
       10000 & 1000 & 2000 & 14 & 0.1  \\
       9000 & 1000 & 2000 & 13 &  0.1\\
       8000 & 1000 & 1000 & 12 & 0.2 \\
       7000 & 1000 & 1000 & 12 & 0.2 \\
       6000 & 1000 & 1000 & 12 & 0.2 \\
       5000 & 1000 & 1000 & 12 & 0.35 \\
       4000 & 1000 & 1000 & 12 & 0.6 \\
       3000 & 600 & 600 & 10 & 0.5 \\
       2000 & 400 & 400 & 9 & 0.5 \\
       1000 & 200 & 200 & 7 & 0.6 \\
       500 & 100 & 100 & 5 & 0.6  \\
       100 & 20 & 20 & 3 & 0.6  \\\hline\hline
       $p$ & \multicolumn{2}{c|}{learning rate}   & $N_{\rm sample}$ &  $N_{\rm epochs}$\\\hline 
      0.1&  \multicolumn{2}{c|}{$10^{-3}$ (constant)}  & $\infty$ & 60000 \\
    \end{tabular}
    \caption{Model parameters of the RNNs for the $L=8$, $p=0.1$ circuit with various dataset sizes $M$. For detailed explanations of these parameters, refer to the main text. }
    \label{tab:vrnn_params}
\end{table}
\begin{table}[t]
    \centering
    \begin{tabular}{c|c|c|c|c|c|c}
       $M$ & batch size & val. size &  $N_h$  & $p_{\rm dropout}$ & $N_{\rm epochs}$ & $N_{\rm sample}$ \\\hline
       12000 & 1000 & 2000 & 19 & 0.1 & 40000 & 20000\\
       10000 & 1000 & 2000 & 18 & 0.1 & 40000 & 20000 \\
       7000 & 1000 & 1000 & 18 & 0.2 & 40000 & 50000 \\
       5000 & 1000 & 1000 & 17 & 0.2 & 40000 & 20000\\
       3000 & 600 & 600 & 14 & 0.2 & 40000 & 50000 \\
       1000 & 200 & 200 & 12 & 0.4 & 60000 & 50000\\
       200 & 40 & 40 & 6 & 0.5  & 60000 & 50000 \\
    \end{tabular}
    \caption{Model parameters of the RNNs for the $L=8$, $p=0.2$ circuit with various dataset sizes $M$. For detailed explanations of these parameters, refer to the main text. }
    \label{tab:vrnn_params-p02}
\end{table}

\bibliography{reference}

\begin{thebibliography}{78}%
\makeatletter
\providecommand \@ifxundefined [1]{%
 \@ifx{#1\undefined}
}%
\providecommand \@ifnum [1]{%
 \ifnum #1\expandafter \@firstoftwo
 \else \expandafter \@secondoftwo
 \fi
}%
\providecommand \@ifx [1]{%
 \ifx #1\expandafter \@firstoftwo
 \else \expandafter \@secondoftwo
 \fi
}%
\providecommand \natexlab [1]{#1}%
\providecommand \enquote  [1]{``#1''}%
\providecommand \bibnamefont  [1]{#1}%
\providecommand \bibfnamefont [1]{#1}%
\providecommand \citenamefont [1]{#1}%
\providecommand \href@noop [0]{\@secondoftwo}%
\providecommand \href [0]{\begingroup \@sanitize@url \@href}%
\providecommand \@href[1]{\@@startlink{#1}\@@href}%
\providecommand \@@href[1]{\endgroup#1\@@endlink}%
\providecommand \@sanitize@url [0]{\catcode `\\12\catcode `\$12\catcode `\&12\catcode `\#12\catcode `\^12\catcode `\_12\catcode `\%12\relax}%
\providecommand \@@startlink[1]{}%
\providecommand \@@endlink[0]{}%
\providecommand \url  [0]{\begingroup\@sanitize@url \@url }%
\providecommand \@url [1]{\endgroup\@href {#1}{\urlprefix }}%
\providecommand \urlprefix  [0]{URL }%
\providecommand \Eprint [0]{\href }%
\providecommand \doibase [0]{https://doi.org/}%
\providecommand \selectlanguage [0]{\@gobble}%
\providecommand \bibinfo  [0]{\@secondoftwo}%
\providecommand \bibfield  [0]{\@secondoftwo}%
\providecommand \translation [1]{[#1]}%
\providecommand \BibitemOpen [0]{}%
\providecommand \bibitemStop [0]{}%
\providecommand \bibitemNoStop [0]{.\EOS\space}%
\providecommand \EOS [0]{\spacefactor3000\relax}%
\providecommand \BibitemShut  [1]{\csname bibitem#1\endcsname}%
\let\auto@bib@innerbib\@empty
\bibitem [{\citenamefont {Kaplan}\ \emph {et~al.}(2020)\citenamefont {Kaplan}, \citenamefont {McCandlish}, \citenamefont {Henighan}, \citenamefont {Brown}, \citenamefont {Chess}, \citenamefont {Child}, \citenamefont {Gray}, \citenamefont {Radford}, \citenamefont {Wu},\ and\ \citenamefont {Amodei}}]{Kaplan:2020trn}%
  \BibitemOpen
  \bibfield  {author} {\bibinfo {author} {\bibfnamefont {J.}~\bibnamefont {Kaplan}}, \bibinfo {author} {\bibfnamefont {S.}~\bibnamefont {McCandlish}}, \bibinfo {author} {\bibfnamefont {T.}~\bibnamefont {Henighan}}, \bibinfo {author} {\bibfnamefont {T.~B.}\ \bibnamefont {Brown}}, \bibinfo {author} {\bibfnamefont {B.}~\bibnamefont {Chess}}, \bibinfo {author} {\bibfnamefont {R.}~\bibnamefont {Child}}, \bibinfo {author} {\bibfnamefont {S.}~\bibnamefont {Gray}}, \bibinfo {author} {\bibfnamefont {A.}~\bibnamefont {Radford}}, \bibinfo {author} {\bibfnamefont {J.}~\bibnamefont {Wu}},\ and\ \bibinfo {author} {\bibfnamefont {D.}~\bibnamefont {Amodei}},\ }\bibfield  {title} {\bibinfo {title} {{Scaling Laws for Neural Language Models}},\ }\href@noop {} {\  (\bibinfo {year} {2020})},\ \Eprint {https://arxiv.org/abs/2001.08361} {arXiv:2001.08361 [cs.LG]} \BibitemShut {NoStop}%
\bibitem [{\citenamefont {Wei}\ \emph {et~al.}(2022)\citenamefont {Wei}, \citenamefont {Tay}, \citenamefont {Bommasani}, \citenamefont {Raffel}, \citenamefont {Zoph}, \citenamefont {Borgeaud}, \citenamefont {Yogatama}, \citenamefont {Bosma}, \citenamefont {Zhou}, \citenamefont {Metzler} \emph {et~al.}}]{wei2022emergent}%
  \BibitemOpen
  \bibfield  {author} {\bibinfo {author} {\bibfnamefont {J.}~\bibnamefont {Wei}}, \bibinfo {author} {\bibfnamefont {Y.}~\bibnamefont {Tay}}, \bibinfo {author} {\bibfnamefont {R.}~\bibnamefont {Bommasani}}, \bibinfo {author} {\bibfnamefont {C.}~\bibnamefont {Raffel}}, \bibinfo {author} {\bibfnamefont {B.}~\bibnamefont {Zoph}}, \bibinfo {author} {\bibfnamefont {S.}~\bibnamefont {Borgeaud}}, \bibinfo {author} {\bibfnamefont {D.}~\bibnamefont {Yogatama}}, \bibinfo {author} {\bibfnamefont {M.}~\bibnamefont {Bosma}}, \bibinfo {author} {\bibfnamefont {D.}~\bibnamefont {Zhou}}, \bibinfo {author} {\bibfnamefont {D.}~\bibnamefont {Metzler}}, \emph {et~al.},\ }\bibfield  {title} {\bibinfo {title} {Emergent abilities of large language models},\ }\href@noop {} {\bibfield  {journal} {\bibinfo  {journal} {arXiv preprint arXiv:2206.07682}\ } (\bibinfo {year} {2022})}\BibitemShut {NoStop}%
\bibitem [{\citenamefont {Carrasquilla}\ \emph {et~al.}(2019{\natexlab{a}})\citenamefont {Carrasquilla}, \citenamefont {Torlai}, \citenamefont {Melko},\ and\ \citenamefont {Aolita}}]{carrasquilla2019reconstructing}%
  \BibitemOpen
  \bibfield  {author} {\bibinfo {author} {\bibfnamefont {J.}~\bibnamefont {Carrasquilla}}, \bibinfo {author} {\bibfnamefont {G.}~\bibnamefont {Torlai}}, \bibinfo {author} {\bibfnamefont {R.~G.}\ \bibnamefont {Melko}},\ and\ \bibinfo {author} {\bibfnamefont {L.}~\bibnamefont {Aolita}},\ }\bibfield  {title} {\bibinfo {title} {Reconstructing quantum states with generative models},\ }\href@noop {} {\bibfield  {journal} {\bibinfo  {journal} {Nature Machine Intelligence}\ }\textbf {\bibinfo {volume} {1}},\ \bibinfo {pages} {155} (\bibinfo {year} {2019}{\natexlab{a}})}\BibitemShut {NoStop}%
\bibitem [{\citenamefont {Torlai}\ and\ \citenamefont {Melko}(2020)}]{torlai2020machine}%
  \BibitemOpen
  \bibfield  {author} {\bibinfo {author} {\bibfnamefont {G.}~\bibnamefont {Torlai}}\ and\ \bibinfo {author} {\bibfnamefont {R.~G.}\ \bibnamefont {Melko}},\ }\bibfield  {title} {\bibinfo {title} {Machine-learning quantum states in the nisq era},\ }\href@noop {} {\bibfield  {journal} {\bibinfo  {journal} {Annual Review of Condensed Matter Physics}\ }\textbf {\bibinfo {volume} {11}},\ \bibinfo {pages} {325} (\bibinfo {year} {2020})}\BibitemShut {NoStop}%
\bibitem [{\citenamefont {Sweke}\ \emph {et~al.}(2020)\citenamefont {Sweke}, \citenamefont {Kesselring}, \citenamefont {van Nieuwenburg},\ and\ \citenamefont {Eisert}}]{sweke2020reinforcement}%
  \BibitemOpen
  \bibfield  {author} {\bibinfo {author} {\bibfnamefont {R.}~\bibnamefont {Sweke}}, \bibinfo {author} {\bibfnamefont {M.~S.}\ \bibnamefont {Kesselring}}, \bibinfo {author} {\bibfnamefont {E.~P.}\ \bibnamefont {van Nieuwenburg}},\ and\ \bibinfo {author} {\bibfnamefont {J.}~\bibnamefont {Eisert}},\ }\bibfield  {title} {\bibinfo {title} {Reinforcement learning decoders for fault-tolerant quantum computation},\ }\href@noop {} {\bibfield  {journal} {\bibinfo  {journal} {Machine Learning: Science and Technology}\ }\textbf {\bibinfo {volume} {2}},\ \bibinfo {pages} {025005} (\bibinfo {year} {2020})}\BibitemShut {NoStop}%
\bibitem [{\citenamefont {Durrer}\ \emph {et~al.}(2020)\citenamefont {Durrer}, \citenamefont {Kratochwil}, \citenamefont {Koski}, \citenamefont {Landig}, \citenamefont {Reichl}, \citenamefont {Wegscheider}, \citenamefont {Ihn},\ and\ \citenamefont {Greplova}}]{durrer2020automated}%
  \BibitemOpen
  \bibfield  {author} {\bibinfo {author} {\bibfnamefont {R.}~\bibnamefont {Durrer}}, \bibinfo {author} {\bibfnamefont {B.}~\bibnamefont {Kratochwil}}, \bibinfo {author} {\bibfnamefont {J.~V.}\ \bibnamefont {Koski}}, \bibinfo {author} {\bibfnamefont {A.~J.}\ \bibnamefont {Landig}}, \bibinfo {author} {\bibfnamefont {C.}~\bibnamefont {Reichl}}, \bibinfo {author} {\bibfnamefont {W.}~\bibnamefont {Wegscheider}}, \bibinfo {author} {\bibfnamefont {T.}~\bibnamefont {Ihn}},\ and\ \bibinfo {author} {\bibfnamefont {E.}~\bibnamefont {Greplova}},\ }\bibfield  {title} {\bibinfo {title} {Automated tuning of double quantum dots into specific charge states using neural networks},\ }\href@noop {} {\bibfield  {journal} {\bibinfo  {journal} {Physical Review Applied}\ }\textbf {\bibinfo {volume} {13}},\ \bibinfo {pages} {054019} (\bibinfo {year} {2020})}\BibitemShut {NoStop}%
\bibitem [{\citenamefont {Moon}\ \emph {et~al.}(2020)\citenamefont {Moon}, \citenamefont {Lennon}, \citenamefont {Kirkpatrick}, \citenamefont {van Esbroeck}, \citenamefont {Camenzind}, \citenamefont {Yu}, \citenamefont {Vigneau}, \citenamefont {Zumb{\"u}hl}, \citenamefont {Briggs}, \citenamefont {Osborne} \emph {et~al.}}]{moon2020machine}%
  \BibitemOpen
  \bibfield  {author} {\bibinfo {author} {\bibfnamefont {H.}~\bibnamefont {Moon}}, \bibinfo {author} {\bibfnamefont {D.~T.}\ \bibnamefont {Lennon}}, \bibinfo {author} {\bibfnamefont {J.}~\bibnamefont {Kirkpatrick}}, \bibinfo {author} {\bibfnamefont {N.~M.}\ \bibnamefont {van Esbroeck}}, \bibinfo {author} {\bibfnamefont {L.~C.}\ \bibnamefont {Camenzind}}, \bibinfo {author} {\bibfnamefont {L.}~\bibnamefont {Yu}}, \bibinfo {author} {\bibfnamefont {F.}~\bibnamefont {Vigneau}}, \bibinfo {author} {\bibfnamefont {D.~M.}\ \bibnamefont {Zumb{\"u}hl}}, \bibinfo {author} {\bibfnamefont {G.~A.~D.}\ \bibnamefont {Briggs}}, \bibinfo {author} {\bibfnamefont {M.~A.}\ \bibnamefont {Osborne}}, \emph {et~al.},\ }\bibfield  {title} {\bibinfo {title} {Machine learning enables completely automatic tuning of a quantum device faster than human experts},\ }\href@noop {} {\bibfield  {journal} {\bibinfo  {journal} {Nature communications}\ }\textbf {\bibinfo {volume} {11}},\ \bibinfo {pages} {4161} (\bibinfo {year} {2020})}\BibitemShut
  {NoStop}%
\bibitem [{\citenamefont {Teoh}\ \emph {et~al.}(2020)\citenamefont {Teoh}, \citenamefont {Drygala}, \citenamefont {Melko},\ and\ \citenamefont {Islam}}]{teoh2020machine}%
  \BibitemOpen
  \bibfield  {author} {\bibinfo {author} {\bibfnamefont {Y.~H.}\ \bibnamefont {Teoh}}, \bibinfo {author} {\bibfnamefont {M.}~\bibnamefont {Drygala}}, \bibinfo {author} {\bibfnamefont {R.~G.}\ \bibnamefont {Melko}},\ and\ \bibinfo {author} {\bibfnamefont {R.}~\bibnamefont {Islam}},\ }\bibfield  {title} {\bibinfo {title} {Machine learning design of a trapped-ion quantum spin simulator},\ }\href@noop {} {\bibfield  {journal} {\bibinfo  {journal} {Quantum Science and Technology}\ }\textbf {\bibinfo {volume} {5}},\ \bibinfo {pages} {024001} (\bibinfo {year} {2020})}\BibitemShut {NoStop}%
\bibitem [{\citenamefont {Carrasquilla}(2020)}]{carrasquilla2020machine}%
  \BibitemOpen
  \bibfield  {author} {\bibinfo {author} {\bibfnamefont {J.}~\bibnamefont {Carrasquilla}},\ }\bibfield  {title} {\bibinfo {title} {Machine learning for quantum matter},\ }\href@noop {} {\bibfield  {journal} {\bibinfo  {journal} {Advances in Physics: X}\ }\textbf {\bibinfo {volume} {5}},\ \bibinfo {pages} {1797528} (\bibinfo {year} {2020})}\BibitemShut {NoStop}%
\bibitem [{\citenamefont {Czischek}\ \emph {et~al.}(2021{\natexlab{a}})\citenamefont {Czischek}, \citenamefont {Yon}, \citenamefont {Genest}, \citenamefont {Roux}, \citenamefont {Rochette}, \citenamefont {Lemyre}, \citenamefont {Moras}, \citenamefont {Pioro-Ladri{\`e}re}, \citenamefont {Drouin}, \citenamefont {Beilliard} \emph {et~al.}}]{czischek2021miniaturizing}%
  \BibitemOpen
  \bibfield  {author} {\bibinfo {author} {\bibfnamefont {S.}~\bibnamefont {Czischek}}, \bibinfo {author} {\bibfnamefont {V.}~\bibnamefont {Yon}}, \bibinfo {author} {\bibfnamefont {M.-A.}\ \bibnamefont {Genest}}, \bibinfo {author} {\bibfnamefont {M.-A.}\ \bibnamefont {Roux}}, \bibinfo {author} {\bibfnamefont {S.}~\bibnamefont {Rochette}}, \bibinfo {author} {\bibfnamefont {J.~C.}\ \bibnamefont {Lemyre}}, \bibinfo {author} {\bibfnamefont {M.}~\bibnamefont {Moras}}, \bibinfo {author} {\bibfnamefont {M.}~\bibnamefont {Pioro-Ladri{\`e}re}}, \bibinfo {author} {\bibfnamefont {D.}~\bibnamefont {Drouin}}, \bibinfo {author} {\bibfnamefont {Y.}~\bibnamefont {Beilliard}}, \emph {et~al.},\ }\bibfield  {title} {\bibinfo {title} {Miniaturizing neural networks for charge state autotuning in quantum dots},\ }\href@noop {} {\bibfield  {journal} {\bibinfo  {journal} {Machine Learning: Science and Technology}\ }\textbf {\bibinfo {volume} {3}},\ \bibinfo {pages} {015001} (\bibinfo {year} {2021}{\natexlab{a}})}\BibitemShut
  {NoStop}%
\bibitem [{\citenamefont {Dawid}\ \emph {et~al.}(2022)\citenamefont {Dawid} \emph {et~al.}}]{Dawid:2022fga}%
  \BibitemOpen
  \bibfield  {author} {\bibinfo {author} {\bibfnamefont {A.}~\bibnamefont {Dawid}} \emph {et~al.},\ }\bibfield  {title} {\bibinfo {title} {{Modern applications of machine learning in quantum sciences}}\ }(\bibinfo {year} {2022})\ \Eprint {https://arxiv.org/abs/2204.04198} {arXiv:2204.04198 [quant-ph]} \BibitemShut {NoStop}%
\bibitem [{\citenamefont {Torlai}\ \emph {et~al.}(2023)\citenamefont {Torlai}, \citenamefont {Wood}, \citenamefont {Acharya}, \citenamefont {Carleo}, \citenamefont {Carrasquilla},\ and\ \citenamefont {Aolita}}]{torlai2023quantum}%
  \BibitemOpen
  \bibfield  {author} {\bibinfo {author} {\bibfnamefont {G.}~\bibnamefont {Torlai}}, \bibinfo {author} {\bibfnamefont {C.~J.}\ \bibnamefont {Wood}}, \bibinfo {author} {\bibfnamefont {A.}~\bibnamefont {Acharya}}, \bibinfo {author} {\bibfnamefont {G.}~\bibnamefont {Carleo}}, \bibinfo {author} {\bibfnamefont {J.}~\bibnamefont {Carrasquilla}},\ and\ \bibinfo {author} {\bibfnamefont {L.}~\bibnamefont {Aolita}},\ }\bibfield  {title} {\bibinfo {title} {Quantum process tomography with unsupervised learning and tensor networks},\ }\href@noop {} {\bibfield  {journal} {\bibinfo  {journal} {Nature Communications}\ }\textbf {\bibinfo {volume} {14}},\ \bibinfo {pages} {2858} (\bibinfo {year} {2023})}\BibitemShut {NoStop}%
\bibitem [{\citenamefont {Zwolak}\ and\ \citenamefont {Taylor}(2023)}]{zwolak2023colloquium}%
  \BibitemOpen
  \bibfield  {author} {\bibinfo {author} {\bibfnamefont {J.~P.}\ \bibnamefont {Zwolak}}\ and\ \bibinfo {author} {\bibfnamefont {J.~M.}\ \bibnamefont {Taylor}},\ }\bibfield  {title} {\bibinfo {title} {Colloquium: Advances in automation of quantum dot devices control},\ }\href@noop {} {\bibfield  {journal} {\bibinfo  {journal} {Reviews of modern physics}\ }\textbf {\bibinfo {volume} {95}},\ \bibinfo {pages} {011006} (\bibinfo {year} {2023})}\BibitemShut {NoStop}%
\bibitem [{\citenamefont {Melko}\ and\ \citenamefont {Carrasquilla}(2024)}]{Melko2024-ev}%
  \BibitemOpen
  \bibfield  {author} {\bibinfo {author} {\bibfnamefont {R.~G.}\ \bibnamefont {Melko}}\ and\ \bibinfo {author} {\bibfnamefont {J.}~\bibnamefont {Carrasquilla}},\ }\bibfield  {title} {\bibinfo {title} {Language models for quantum simulation},\ }\href {https://doi.org/10.1038/s43588-023-00578-0} {\bibfield  {journal} {\bibinfo  {journal} {Nature Computational Science}\ }\textbf {\bibinfo {volume} {4}},\ \bibinfo {pages} {11} (\bibinfo {year} {2024})}\BibitemShut {NoStop}%
\bibitem [{\citenamefont {Li}\ \emph {et~al.}(2023{\natexlab{a}})\citenamefont {Li}, \citenamefont {Zou}, \citenamefont {Glorioso}, \citenamefont {Altman},\ and\ \citenamefont {Fisher}}]{Li:2022vks}%
  \BibitemOpen
  \bibfield  {author} {\bibinfo {author} {\bibfnamefont {Y.}~\bibnamefont {Li}}, \bibinfo {author} {\bibfnamefont {Y.}~\bibnamefont {Zou}}, \bibinfo {author} {\bibfnamefont {P.}~\bibnamefont {Glorioso}}, \bibinfo {author} {\bibfnamefont {E.}~\bibnamefont {Altman}},\ and\ \bibinfo {author} {\bibfnamefont {M.~P.~A.}\ \bibnamefont {Fisher}},\ }\bibfield  {title} {\bibinfo {title} {{Cross Entropy Benchmark for Measurement-Induced Phase Transitions}},\ }\href {https://doi.org/10.1103/PhysRevLett.130.220404} {\bibfield  {journal} {\bibinfo  {journal} {Phys. Rev. Lett.}\ }\textbf {\bibinfo {volume} {130}},\ \bibinfo {pages} {220404} (\bibinfo {year} {2023}{\natexlab{a}})},\ \Eprint {https://arxiv.org/abs/2209.00609} {arXiv:2209.00609 [quant-ph]} \BibitemShut {NoStop}%
\bibitem [{\citenamefont {Skinner}\ \emph {et~al.}(2019)\citenamefont {Skinner}, \citenamefont {Ruhman},\ and\ \citenamefont {Nahum}}]{Skinner:2018tjl}%
  \BibitemOpen
  \bibfield  {author} {\bibinfo {author} {\bibfnamefont {B.}~\bibnamefont {Skinner}}, \bibinfo {author} {\bibfnamefont {J.}~\bibnamefont {Ruhman}},\ and\ \bibinfo {author} {\bibfnamefont {A.}~\bibnamefont {Nahum}},\ }\bibfield  {title} {\bibinfo {title} {{Measurement-Induced Phase Transitions in the Dynamics of Entanglement}},\ }\href {https://doi.org/10.1103/PhysRevX.9.031009} {\bibfield  {journal} {\bibinfo  {journal} {Phys. Rev. X}\ }\textbf {\bibinfo {volume} {9}},\ \bibinfo {pages} {031009} (\bibinfo {year} {2019})},\ \Eprint {https://arxiv.org/abs/1808.05953} {arXiv:1808.05953 [cond-mat.stat-mech]} \BibitemShut {NoStop}%
\bibitem [{\citenamefont {Li}\ \emph {et~al.}(2018)\citenamefont {Li}, \citenamefont {Chen},\ and\ \citenamefont {Fisher}}]{PhysRevB.98.205136}%
  \BibitemOpen
  \bibfield  {author} {\bibinfo {author} {\bibfnamefont {Y.}~\bibnamefont {Li}}, \bibinfo {author} {\bibfnamefont {X.}~\bibnamefont {Chen}},\ and\ \bibinfo {author} {\bibfnamefont {M.~P.~A.}\ \bibnamefont {Fisher}},\ }\bibfield  {title} {\bibinfo {title} {Quantum zeno effect and the many-body entanglement transition},\ }\href {https://doi.org/10.1103/PhysRevB.98.205136} {\bibfield  {journal} {\bibinfo  {journal} {Phys. Rev. B}\ }\textbf {\bibinfo {volume} {98}},\ \bibinfo {pages} {205136} (\bibinfo {year} {2018})}\BibitemShut {NoStop}%
\bibitem [{\citenamefont {Chan}\ \emph {et~al.}(2019)\citenamefont {Chan}, \citenamefont {Nandkishore}, \citenamefont {Pretko},\ and\ \citenamefont {Smith}}]{PhysRevB.99.224307}%
  \BibitemOpen
  \bibfield  {author} {\bibinfo {author} {\bibfnamefont {A.}~\bibnamefont {Chan}}, \bibinfo {author} {\bibfnamefont {R.~M.}\ \bibnamefont {Nandkishore}}, \bibinfo {author} {\bibfnamefont {M.}~\bibnamefont {Pretko}},\ and\ \bibinfo {author} {\bibfnamefont {G.}~\bibnamefont {Smith}},\ }\bibfield  {title} {\bibinfo {title} {Unitary-projective entanglement dynamics},\ }\href {https://doi.org/10.1103/PhysRevB.99.224307} {\bibfield  {journal} {\bibinfo  {journal} {Phys. Rev. B}\ }\textbf {\bibinfo {volume} {99}},\ \bibinfo {pages} {224307} (\bibinfo {year} {2019})}\BibitemShut {NoStop}%
\bibitem [{\citenamefont {Choi}\ \emph {et~al.}(2020)\citenamefont {Choi}, \citenamefont {Bao}, \citenamefont {Qi},\ and\ \citenamefont {Altman}}]{Choi:2019nhg}%
  \BibitemOpen
  \bibfield  {author} {\bibinfo {author} {\bibfnamefont {S.}~\bibnamefont {Choi}}, \bibinfo {author} {\bibfnamefont {Y.}~\bibnamefont {Bao}}, \bibinfo {author} {\bibfnamefont {X.-L.}\ \bibnamefont {Qi}},\ and\ \bibinfo {author} {\bibfnamefont {E.}~\bibnamefont {Altman}},\ }\bibfield  {title} {\bibinfo {title} {{Quantum Error Correction in Scrambling Dynamics and Measurement-Induced Phase Transition}},\ }\href {https://doi.org/10.1103/PhysRevLett.125.030505} {\bibfield  {journal} {\bibinfo  {journal} {Phys. Rev. Lett.}\ }\textbf {\bibinfo {volume} {125}},\ \bibinfo {pages} {030505} (\bibinfo {year} {2020})},\ \Eprint {https://arxiv.org/abs/1903.05124} {arXiv:1903.05124 [quant-ph]} \BibitemShut {NoStop}%
\bibitem [{\citenamefont {Jian}\ \emph {et~al.}(2020{\natexlab{a}})\citenamefont {Jian}, \citenamefont {You}, \citenamefont {Vasseur},\ and\ \citenamefont {Ludwig}}]{Jian:2019mny}%
  \BibitemOpen
  \bibfield  {author} {\bibinfo {author} {\bibfnamefont {C.-M.}\ \bibnamefont {Jian}}, \bibinfo {author} {\bibfnamefont {Y.-Z.}\ \bibnamefont {You}}, \bibinfo {author} {\bibfnamefont {R.}~\bibnamefont {Vasseur}},\ and\ \bibinfo {author} {\bibfnamefont {A.~W.~W.}\ \bibnamefont {Ludwig}},\ }\bibfield  {title} {\bibinfo {title} {{Measurement-induced criticality in random quantum circuits}},\ }\href {https://doi.org/10.1103/PhysRevB.101.104302} {\bibfield  {journal} {\bibinfo  {journal} {Phys. Rev. B}\ }\textbf {\bibinfo {volume} {101}},\ \bibinfo {pages} {104302} (\bibinfo {year} {2020}{\natexlab{a}})},\ \Eprint {https://arxiv.org/abs/1908.08051} {arXiv:1908.08051 [cond-mat.stat-mech]} \BibitemShut {NoStop}%
\bibitem [{\citenamefont {Li}\ \emph {et~al.}(2019{\natexlab{a}})\citenamefont {Li}, \citenamefont {Chen},\ and\ \citenamefont {Fisher}}]{Li_2019}%
  \BibitemOpen
  \bibfield  {author} {\bibinfo {author} {\bibfnamefont {Y.}~\bibnamefont {Li}}, \bibinfo {author} {\bibfnamefont {X.}~\bibnamefont {Chen}},\ and\ \bibinfo {author} {\bibfnamefont {M.~P.~A.}\ \bibnamefont {Fisher}},\ }\bibfield  {title} {\bibinfo {title} {Measurement-driven entanglement transition in hybrid quantum circuits},\ }\href {https://doi.org/10.1103/PhysRevB.100.134306} {\bibfield  {journal} {\bibinfo  {journal} {Phys. Rev. B}\ }\textbf {\bibinfo {volume} {100}},\ \bibinfo {pages} {134306} (\bibinfo {year} {2019}{\natexlab{a}})}\BibitemShut {NoStop}%
\bibitem [{\citenamefont {Zabalo}\ \emph {et~al.}(2020)\citenamefont {Zabalo}, \citenamefont {Gullans}, \citenamefont {Wilson}, \citenamefont {Gopalakrishnan}, \citenamefont {Huse},\ and\ \citenamefont {Pixley}}]{Zabalo:2019sfl}%
  \BibitemOpen
  \bibfield  {author} {\bibinfo {author} {\bibfnamefont {A.}~\bibnamefont {Zabalo}}, \bibinfo {author} {\bibfnamefont {M.~J.}\ \bibnamefont {Gullans}}, \bibinfo {author} {\bibfnamefont {J.~H.}\ \bibnamefont {Wilson}}, \bibinfo {author} {\bibfnamefont {S.}~\bibnamefont {Gopalakrishnan}}, \bibinfo {author} {\bibfnamefont {D.~A.}\ \bibnamefont {Huse}},\ and\ \bibinfo {author} {\bibfnamefont {J.~H.}\ \bibnamefont {Pixley}},\ }\bibfield  {title} {\bibinfo {title} {{Critical properties of the measurement-induced transition in random quantum circuits}},\ }\href {https://doi.org/10.1103/PhysRevB.101.060301} {\bibfield  {journal} {\bibinfo  {journal} {Phys. Rev. B}\ }\textbf {\bibinfo {volume} {101}},\ \bibinfo {pages} {060301} (\bibinfo {year} {2020})},\ \Eprint {https://arxiv.org/abs/1911.00008} {arXiv:1911.00008 [cond-mat.dis-nn]} \BibitemShut {NoStop}%
\bibitem [{\citenamefont {Bao}\ \emph {et~al.}(2020)\citenamefont {Bao}, \citenamefont {Choi},\ and\ \citenamefont {Altman}}]{Bao_2020}%
  \BibitemOpen
  \bibfield  {author} {\bibinfo {author} {\bibfnamefont {Y.}~\bibnamefont {Bao}}, \bibinfo {author} {\bibfnamefont {S.}~\bibnamefont {Choi}},\ and\ \bibinfo {author} {\bibfnamefont {E.}~\bibnamefont {Altman}},\ }\bibfield  {title} {\bibinfo {title} {Theory of the phase transition in random unitary circuits with measurements},\ }\href {https://doi.org/10.1103/PhysRevB.101.104301} {\bibfield  {journal} {\bibinfo  {journal} {Phys. Rev. B}\ }\textbf {\bibinfo {volume} {101}},\ \bibinfo {pages} {104301} (\bibinfo {year} {2020})}\BibitemShut {NoStop}%
\bibitem [{\citenamefont {Gullans}\ and\ \citenamefont {Huse}(2020)}]{PhysRevX.10.041020}%
  \BibitemOpen
  \bibfield  {author} {\bibinfo {author} {\bibfnamefont {M.~J.}\ \bibnamefont {Gullans}}\ and\ \bibinfo {author} {\bibfnamefont {D.~A.}\ \bibnamefont {Huse}},\ }\bibfield  {title} {\bibinfo {title} {Dynamical purification phase transition induced by quantum measurements},\ }\href {https://doi.org/10.1103/PhysRevX.10.041020} {\bibfield  {journal} {\bibinfo  {journal} {Phys. Rev. X}\ }\textbf {\bibinfo {volume} {10}},\ \bibinfo {pages} {041020} (\bibinfo {year} {2020})}\BibitemShut {NoStop}%
\bibitem [{\citenamefont {Jian}\ \emph {et~al.}(2020{\natexlab{b}})\citenamefont {Jian}, \citenamefont {You}, \citenamefont {Vasseur},\ and\ \citenamefont {Ludwig}}]{PhysRevB.101.104302}%
  \BibitemOpen
  \bibfield  {author} {\bibinfo {author} {\bibfnamefont {C.-M.}\ \bibnamefont {Jian}}, \bibinfo {author} {\bibfnamefont {Y.-Z.}\ \bibnamefont {You}}, \bibinfo {author} {\bibfnamefont {R.}~\bibnamefont {Vasseur}},\ and\ \bibinfo {author} {\bibfnamefont {A.~W.~W.}\ \bibnamefont {Ludwig}},\ }\bibfield  {title} {\bibinfo {title} {Measurement-induced criticality in random quantum circuits},\ }\href {https://doi.org/10.1103/PhysRevB.101.104302} {\bibfield  {journal} {\bibinfo  {journal} {Phys. Rev. B}\ }\textbf {\bibinfo {volume} {101}},\ \bibinfo {pages} {104302} (\bibinfo {year} {2020}{\natexlab{b}})}\BibitemShut {NoStop}%
\bibitem [{\citenamefont {Agrawal}\ \emph {et~al.}(2022)\citenamefont {Agrawal}, \citenamefont {Zabalo}, \citenamefont {Chen}, \citenamefont {Wilson}, \citenamefont {Potter}, \citenamefont {Pixley}, \citenamefont {Gopalakrishnan},\ and\ \citenamefont {Vasseur}}]{Agrawal:2021ukw}%
  \BibitemOpen
  \bibfield  {author} {\bibinfo {author} {\bibfnamefont {U.}~\bibnamefont {Agrawal}}, \bibinfo {author} {\bibfnamefont {A.}~\bibnamefont {Zabalo}}, \bibinfo {author} {\bibfnamefont {K.}~\bibnamefont {Chen}}, \bibinfo {author} {\bibfnamefont {J.~H.}\ \bibnamefont {Wilson}}, \bibinfo {author} {\bibfnamefont {A.~C.}\ \bibnamefont {Potter}}, \bibinfo {author} {\bibfnamefont {J.~H.}\ \bibnamefont {Pixley}}, \bibinfo {author} {\bibfnamefont {S.}~\bibnamefont {Gopalakrishnan}},\ and\ \bibinfo {author} {\bibfnamefont {R.}~\bibnamefont {Vasseur}},\ }\bibfield  {title} {\bibinfo {title} {{Entanglement and Charge-Sharpening Transitions in U(1) Symmetric Monitored Quantum Circuits}},\ }\href {https://doi.org/10.1103/PhysRevX.12.041002} {\bibfield  {journal} {\bibinfo  {journal} {Phys. Rev. X}\ }\textbf {\bibinfo {volume} {12}},\ \bibinfo {pages} {041002} (\bibinfo {year} {2022})},\ \Eprint {https://arxiv.org/abs/2107.10279} {arXiv:2107.10279 [cond-mat.dis-nn]} \BibitemShut {NoStop}%
\bibitem [{\citenamefont {Li}\ \emph {et~al.}(2021)\citenamefont {Li}, \citenamefont {Chen}, \citenamefont {Ludwig},\ and\ \citenamefont {Fisher}}]{Li_2021v2}%
  \BibitemOpen
  \bibfield  {author} {\bibinfo {author} {\bibfnamefont {Y.}~\bibnamefont {Li}}, \bibinfo {author} {\bibfnamefont {X.}~\bibnamefont {Chen}}, \bibinfo {author} {\bibfnamefont {A.~W.~W.}\ \bibnamefont {Ludwig}},\ and\ \bibinfo {author} {\bibfnamefont {M.~P.~A.}\ \bibnamefont {Fisher}},\ }\bibfield  {title} {\bibinfo {title} {Conformal invariance and quantum nonlocality in critical hybrid circuits},\ }\href {https://doi.org/10.1103/PhysRevB.104.104305} {\bibfield  {journal} {\bibinfo  {journal} {Phys. Rev. B}\ }\textbf {\bibinfo {volume} {104}},\ \bibinfo {pages} {104305} (\bibinfo {year} {2021})}\BibitemShut {NoStop}%
\bibitem [{\citenamefont {Sang}\ \emph {et~al.}(2021)\citenamefont {Sang}, \citenamefont {Li}, \citenamefont {Zhou}, \citenamefont {Chen}, \citenamefont {Hsieh},\ and\ \citenamefont {Fisher}}]{Sang_2021}%
  \BibitemOpen
  \bibfield  {author} {\bibinfo {author} {\bibfnamefont {S.}~\bibnamefont {Sang}}, \bibinfo {author} {\bibfnamefont {Y.}~\bibnamefont {Li}}, \bibinfo {author} {\bibfnamefont {T.}~\bibnamefont {Zhou}}, \bibinfo {author} {\bibfnamefont {X.}~\bibnamefont {Chen}}, \bibinfo {author} {\bibfnamefont {T.~H.}\ \bibnamefont {Hsieh}},\ and\ \bibinfo {author} {\bibfnamefont {M.~P.}\ \bibnamefont {Fisher}},\ }\bibfield  {title} {\bibinfo {title} {Entanglement negativity at measurement-induced criticality},\ }\href {https://doi.org/10.1103/PRXQuantum.2.030313} {\bibfield  {journal} {\bibinfo  {journal} {PRX Quantum}\ }\textbf {\bibinfo {volume} {2}},\ \bibinfo {pages} {030313} (\bibinfo {year} {2021})}\BibitemShut {NoStop}%
\bibitem [{\citenamefont {Barratt}\ \emph {et~al.}(2022)\citenamefont {Barratt}, \citenamefont {Agrawal}, \citenamefont {Gopalakrishnan}, \citenamefont {Huse}, \citenamefont {Vasseur},\ and\ \citenamefont {Potter}}]{PhysRevLett.129.120604}%
  \BibitemOpen
  \bibfield  {author} {\bibinfo {author} {\bibfnamefont {F.}~\bibnamefont {Barratt}}, \bibinfo {author} {\bibfnamefont {U.}~\bibnamefont {Agrawal}}, \bibinfo {author} {\bibfnamefont {S.}~\bibnamefont {Gopalakrishnan}}, \bibinfo {author} {\bibfnamefont {D.~A.}\ \bibnamefont {Huse}}, \bibinfo {author} {\bibfnamefont {R.}~\bibnamefont {Vasseur}},\ and\ \bibinfo {author} {\bibfnamefont {A.~C.}\ \bibnamefont {Potter}},\ }\bibfield  {title} {\bibinfo {title} {Field theory of charge sharpening in symmetric monitored quantum circuits},\ }\href {https://doi.org/10.1103/PhysRevLett.129.120604} {\bibfield  {journal} {\bibinfo  {journal} {Phys. Rev. Lett.}\ }\textbf {\bibinfo {volume} {129}},\ \bibinfo {pages} {120604} (\bibinfo {year} {2022})}\BibitemShut {NoStop}%
\bibitem [{\citenamefont {Weinstein}\ \emph {et~al.}(2022)\citenamefont {Weinstein}, \citenamefont {Bao},\ and\ \citenamefont {Altman}}]{Weinstein_2022}%
  \BibitemOpen
  \bibfield  {author} {\bibinfo {author} {\bibfnamefont {Z.}~\bibnamefont {Weinstein}}, \bibinfo {author} {\bibfnamefont {Y.}~\bibnamefont {Bao}},\ and\ \bibinfo {author} {\bibfnamefont {E.}~\bibnamefont {Altman}},\ }\bibfield  {title} {\bibinfo {title} {Measurement-induced power-law negativity in an open monitored quantum circuit},\ }\bibfield  {journal} {\bibinfo  {journal} {Physical Review Letters}\ }\textbf {\bibinfo {volume} {129}},\ \href {https://doi.org/10.1103/physrevlett.129.080501} {10.1103/physrevlett.129.080501} (\bibinfo {year} {2022})\BibitemShut {NoStop}%
\bibitem [{\citenamefont {Skinner}(2023)}]{Skinner:2023hle}%
  \BibitemOpen
  \bibfield  {author} {\bibinfo {author} {\bibfnamefont {B.}~\bibnamefont {Skinner}},\ }\bibfield  {title} {\bibinfo {title} {{Lecture Notes: Introduction to random unitary circuits and the measurement-induced entanglement phase transition}},\ }\href@noop {} {\  (\bibinfo {year} {2023})},\ \Eprint {https://arxiv.org/abs/2307.02986} {arXiv:2307.02986 [cond-mat.stat-mech]} \BibitemShut {NoStop}%
\bibitem [{\citenamefont {Li}\ \emph {et~al.}(2023{\natexlab{b}})\citenamefont {Li}, \citenamefont {Vijay},\ and\ \citenamefont {Fisher}}]{Li_KPZ2023}%
  \BibitemOpen
  \bibfield  {author} {\bibinfo {author} {\bibfnamefont {Y.}~\bibnamefont {Li}}, \bibinfo {author} {\bibfnamefont {S.}~\bibnamefont {Vijay}},\ and\ \bibinfo {author} {\bibfnamefont {M.~P.}\ \bibnamefont {Fisher}},\ }\bibfield  {title} {\bibinfo {title} {Entanglement domain walls in monitored quantum circuits and the directed polymer in a random environment},\ }\href {https://doi.org/10.1103/PRXQuantum.4.010331} {\bibfield  {journal} {\bibinfo  {journal} {PRX Quantum}\ }\textbf {\bibinfo {volume} {4}},\ \bibinfo {pages} {010331} (\bibinfo {year} {2023}{\natexlab{b}})}\BibitemShut {NoStop}%
\bibitem [{\citenamefont {Majidy}\ \emph {et~al.}(2023)\citenamefont {Majidy}, \citenamefont {Agrawal}, \citenamefont {Gopalakrishnan}, \citenamefont {Potter}, \citenamefont {Vasseur},\ and\ \citenamefont {Halpern}}]{PhysRevB.108.054307}%
  \BibitemOpen
  \bibfield  {author} {\bibinfo {author} {\bibfnamefont {S.}~\bibnamefont {Majidy}}, \bibinfo {author} {\bibfnamefont {U.}~\bibnamefont {Agrawal}}, \bibinfo {author} {\bibfnamefont {S.}~\bibnamefont {Gopalakrishnan}}, \bibinfo {author} {\bibfnamefont {A.~C.}\ \bibnamefont {Potter}}, \bibinfo {author} {\bibfnamefont {R.}~\bibnamefont {Vasseur}},\ and\ \bibinfo {author} {\bibfnamefont {N.~Y.}\ \bibnamefont {Halpern}},\ }\bibfield  {title} {\bibinfo {title} {Critical phase and spin sharpening in su(2)-symmetric monitored quantum circuits},\ }\href {https://doi.org/10.1103/PhysRevB.108.054307} {\bibfield  {journal} {\bibinfo  {journal} {Phys. Rev. B}\ }\textbf {\bibinfo {volume} {108}},\ \bibinfo {pages} {054307} (\bibinfo {year} {2023})}\BibitemShut {NoStop}%
\bibitem [{\citenamefont {Avakian}\ \emph {et~al.}(2024)\citenamefont {Avakian}, \citenamefont {Pereg-Barnea},\ and\ \citenamefont {Witczak-Krempa}}]{avakian2024longrangemultipartiteentanglementnear}%
  \BibitemOpen
  \bibfield  {author} {\bibinfo {author} {\bibfnamefont {S.~J.}\ \bibnamefont {Avakian}}, \bibinfo {author} {\bibfnamefont {T.}~\bibnamefont {Pereg-Barnea}},\ and\ \bibinfo {author} {\bibfnamefont {W.}~\bibnamefont {Witczak-Krempa}},\ }\href {https://arxiv.org/abs/2404.16095} {\bibinfo {title} {Long-range multipartite entanglement near measurement-induced transitions}} (\bibinfo {year} {2024}),\ \Eprint {https://arxiv.org/abs/2404.16095} {arXiv:2404.16095 [quant-ph]} \BibitemShut {NoStop}%
\bibitem [{\citenamefont {Hu}\ and\ \citenamefont {Zou}(2024)}]{PhysRevB.110.195107}%
  \BibitemOpen
  \bibfield  {author} {\bibinfo {author} {\bibfnamefont {Y.}~\bibnamefont {Hu}}\ and\ \bibinfo {author} {\bibfnamefont {Y.}~\bibnamefont {Zou}},\ }\bibfield  {title} {\bibinfo {title} {Petz map recovery for long-range entangled quantum many-body states},\ }\href {https://doi.org/10.1103/PhysRevB.110.195107} {\bibfield  {journal} {\bibinfo  {journal} {Phys. Rev. B}\ }\textbf {\bibinfo {volume} {110}},\ \bibinfo {pages} {195107} (\bibinfo {year} {2024})}\BibitemShut {NoStop}%
\bibitem [{\citenamefont {Li}\ \emph {et~al.}(2019{\natexlab{b}})\citenamefont {Li}, \citenamefont {Chen},\ and\ \citenamefont {Fisher}}]{PhysRevB.100.134306}%
  \BibitemOpen
  \bibfield  {author} {\bibinfo {author} {\bibfnamefont {Y.}~\bibnamefont {Li}}, \bibinfo {author} {\bibfnamefont {X.}~\bibnamefont {Chen}},\ and\ \bibinfo {author} {\bibfnamefont {M.~P.~A.}\ \bibnamefont {Fisher}},\ }\bibfield  {title} {\bibinfo {title} {Measurement-driven entanglement transition in hybrid quantum circuits},\ }\href {https://doi.org/10.1103/PhysRevB.100.134306} {\bibfield  {journal} {\bibinfo  {journal} {Phys. Rev. B}\ }\textbf {\bibinfo {volume} {100}},\ \bibinfo {pages} {134306} (\bibinfo {year} {2019}{\natexlab{b}})}\BibitemShut {NoStop}%
\bibitem [{\citenamefont {Fisher}\ \emph {et~al.}(2023)\citenamefont {Fisher}, \citenamefont {Khemani}, \citenamefont {Nahum},\ and\ \citenamefont {Vijay}}]{Fisher:2022qey}%
  \BibitemOpen
  \bibfield  {author} {\bibinfo {author} {\bibfnamefont {M.~P.~A.}\ \bibnamefont {Fisher}}, \bibinfo {author} {\bibfnamefont {V.}~\bibnamefont {Khemani}}, \bibinfo {author} {\bibfnamefont {A.}~\bibnamefont {Nahum}},\ and\ \bibinfo {author} {\bibfnamefont {S.}~\bibnamefont {Vijay}},\ }\bibfield  {title} {\bibinfo {title} {{Random Quantum Circuits}},\ }\href {https://doi.org/10.1146/annurev-conmatphys-031720-030658} {\bibfield  {journal} {\bibinfo  {journal} {Ann. Rev. Condensed Matter Phys.}\ }\textbf {\bibinfo {volume} {14}},\ \bibinfo {pages} {335} (\bibinfo {year} {2023})},\ \Eprint {https://arxiv.org/abs/2207.14280} {arXiv:2207.14280 [quant-ph]} \BibitemShut {NoStop}%
\bibitem [{\citenamefont {Agrawal}\ \emph {et~al.}(2024)\citenamefont {Agrawal}, \citenamefont {Lopez-Piqueres}, \citenamefont {Vasseur}, \citenamefont {Gopalakrishnan},\ and\ \citenamefont {Potter}}]{Agrawal:2023kuy}%
  \BibitemOpen
  \bibfield  {author} {\bibinfo {author} {\bibfnamefont {U.}~\bibnamefont {Agrawal}}, \bibinfo {author} {\bibfnamefont {J.}~\bibnamefont {Lopez-Piqueres}}, \bibinfo {author} {\bibfnamefont {R.}~\bibnamefont {Vasseur}}, \bibinfo {author} {\bibfnamefont {S.}~\bibnamefont {Gopalakrishnan}},\ and\ \bibinfo {author} {\bibfnamefont {A.~C.}\ \bibnamefont {Potter}},\ }\bibfield  {title} {\bibinfo {title} {{Observing Quantum Measurement Collapse as a Learnability Phase Transition}},\ }\href {https://doi.org/10.1103/PhysRevX.14.041012} {\bibfield  {journal} {\bibinfo  {journal} {Phys. Rev. X}\ }\textbf {\bibinfo {volume} {14}},\ \bibinfo {pages} {041012} (\bibinfo {year} {2024})},\ \Eprint {https://arxiv.org/abs/2311.00058} {arXiv:2311.00058 [quant-ph]} \BibitemShut {NoStop}%
\bibitem [{\citenamefont {Porras}\ and\ \citenamefont {Cirac}(2004)}]{PhysRevLett.92.207901}%
  \BibitemOpen
  \bibfield  {author} {\bibinfo {author} {\bibfnamefont {D.}~\bibnamefont {Porras}}\ and\ \bibinfo {author} {\bibfnamefont {J.~I.}\ \bibnamefont {Cirac}},\ }\bibfield  {title} {\bibinfo {title} {Effective quantum spin systems with trapped ions},\ }\href {https://doi.org/10.1103/PhysRevLett.92.207901} {\bibfield  {journal} {\bibinfo  {journal} {Phys. Rev. Lett.}\ }\textbf {\bibinfo {volume} {92}},\ \bibinfo {pages} {207901} (\bibinfo {year} {2004})}\BibitemShut {NoStop}%
\bibitem [{\citenamefont {Deng}\ \emph {et~al.}(2005)\citenamefont {Deng}, \citenamefont {Porras},\ and\ \citenamefont {Cirac}}]{PhysRevA.72.063407}%
  \BibitemOpen
  \bibfield  {author} {\bibinfo {author} {\bibfnamefont {X.-L.}\ \bibnamefont {Deng}}, \bibinfo {author} {\bibfnamefont {D.}~\bibnamefont {Porras}},\ and\ \bibinfo {author} {\bibfnamefont {J.~I.}\ \bibnamefont {Cirac}},\ }\bibfield  {title} {\bibinfo {title} {Effective spin quantum phases in systems of trapped ions},\ }\href {https://doi.org/10.1103/PhysRevA.72.063407} {\bibfield  {journal} {\bibinfo  {journal} {Phys. Rev. A}\ }\textbf {\bibinfo {volume} {72}},\ \bibinfo {pages} {063407} (\bibinfo {year} {2005})}\BibitemShut {NoStop}%
\bibitem [{\citenamefont {Myerson}\ \emph {et~al.}(2008)\citenamefont {Myerson}, \citenamefont {Szwer}, \citenamefont {Webster}, \citenamefont {Allcock}, \citenamefont {Curtis}, \citenamefont {Imreh}, \citenamefont {Sherman}, \citenamefont {Stacey}, \citenamefont {Steane},\ and\ \citenamefont {Lucas}}]{PhysRevLett.100.200502}%
  \BibitemOpen
  \bibfield  {author} {\bibinfo {author} {\bibfnamefont {A.~H.}\ \bibnamefont {Myerson}}, \bibinfo {author} {\bibfnamefont {D.~J.}\ \bibnamefont {Szwer}}, \bibinfo {author} {\bibfnamefont {S.~C.}\ \bibnamefont {Webster}}, \bibinfo {author} {\bibfnamefont {D.~T.~C.}\ \bibnamefont {Allcock}}, \bibinfo {author} {\bibfnamefont {M.~J.}\ \bibnamefont {Curtis}}, \bibinfo {author} {\bibfnamefont {G.}~\bibnamefont {Imreh}}, \bibinfo {author} {\bibfnamefont {J.~A.}\ \bibnamefont {Sherman}}, \bibinfo {author} {\bibfnamefont {D.~N.}\ \bibnamefont {Stacey}}, \bibinfo {author} {\bibfnamefont {A.~M.}\ \bibnamefont {Steane}},\ and\ \bibinfo {author} {\bibfnamefont {D.~M.}\ \bibnamefont {Lucas}},\ }\bibfield  {title} {\bibinfo {title} {High-fidelity readout of trapped-ion qubits},\ }\href {https://doi.org/10.1103/PhysRevLett.100.200502} {\bibfield  {journal} {\bibinfo  {journal} {Phys. Rev. Lett.}\ }\textbf {\bibinfo {volume} {100}},\ \bibinfo {pages} {200502} (\bibinfo {year} {2008})}\BibitemShut {NoStop}%
\bibitem [{\citenamefont {Taylor}\ and\ \citenamefont {Calarco}(2008)}]{PhysRevA.78.062331}%
  \BibitemOpen
  \bibfield  {author} {\bibinfo {author} {\bibfnamefont {J.~M.}\ \bibnamefont {Taylor}}\ and\ \bibinfo {author} {\bibfnamefont {T.}~\bibnamefont {Calarco}},\ }\bibfield  {title} {\bibinfo {title} {Wigner crystals of ions as quantum hard drives},\ }\href {https://doi.org/10.1103/PhysRevA.78.062331} {\bibfield  {journal} {\bibinfo  {journal} {Phys. Rev. A}\ }\textbf {\bibinfo {volume} {78}},\ \bibinfo {pages} {062331} (\bibinfo {year} {2008})}\BibitemShut {NoStop}%
\bibitem [{bla(2012)}]{blatt2012quantum}%
  \BibitemOpen
  \bibfield  {title} {\bibinfo {title} {Quantum simulations with trapped ions},\ }\href@noop {} {\bibfield  {journal} {\bibinfo  {journal} {Nature Physics}\ }\textbf {\bibinfo {volume} {8}},\ \bibinfo {pages} {277} (\bibinfo {year} {2012})}\BibitemShut {NoStop}%
\bibitem [{\citenamefont {Debnath}\ \emph {et~al.}(2016)\citenamefont {Debnath}, \citenamefont {Linke}, \citenamefont {Figgatt}, \citenamefont {Landsman}, \citenamefont {Wright},\ and\ \citenamefont {Monroe}}]{Debnath_2016}%
  \BibitemOpen
  \bibfield  {author} {\bibinfo {author} {\bibfnamefont {S.}~\bibnamefont {Debnath}}, \bibinfo {author} {\bibfnamefont {N.~M.}\ \bibnamefont {Linke}}, \bibinfo {author} {\bibfnamefont {C.}~\bibnamefont {Figgatt}}, \bibinfo {author} {\bibfnamefont {K.~A.}\ \bibnamefont {Landsman}}, \bibinfo {author} {\bibfnamefont {K.}~\bibnamefont {Wright}},\ and\ \bibinfo {author} {\bibfnamefont {C.}~\bibnamefont {Monroe}},\ }\bibfield  {title} {\bibinfo {title} {Demonstration of a small programmable quantum computer with atomic qubits},\ }\href {https://doi.org/10.1038/nature18648} {\bibfield  {journal} {\bibinfo  {journal} {Nature}\ }\textbf {\bibinfo {volume} {536}},\ \bibinfo {pages} {63–66} (\bibinfo {year} {2016})}\BibitemShut {NoStop}%
\bibitem [{\citenamefont {Bruzewicz}\ \emph {et~al.}(2019)\citenamefont {Bruzewicz}, \citenamefont {Chiaverini}, \citenamefont {McConnell},\ and\ \citenamefont {Sage}}]{10.1063/1.5088164}%
  \BibitemOpen
  \bibfield  {author} {\bibinfo {author} {\bibfnamefont {C.~D.}\ \bibnamefont {Bruzewicz}}, \bibinfo {author} {\bibfnamefont {J.}~\bibnamefont {Chiaverini}}, \bibinfo {author} {\bibfnamefont {R.}~\bibnamefont {McConnell}},\ and\ \bibinfo {author} {\bibfnamefont {J.~M.}\ \bibnamefont {Sage}},\ }\bibfield  {title} {\bibinfo {title} {Trapped-ion quantum computing: Progress and challenges},\ }\href {https://doi.org/10.1063/1.5088164} {\bibfield  {journal} {\bibinfo  {journal} {Applied Physics Reviews}\ }\textbf {\bibinfo {volume} {6}},\ \bibinfo {pages} {021314} (\bibinfo {year} {2019})},\ \Eprint {https://arxiv.org/abs/https://pubs.aip.org/aip/apr/article-pdf/doi/10.1063/1.5088164/19742554/021314\_1\_online.pdf} {https://pubs.aip.org/aip/apr/article-pdf/doi/10.1063/1.5088164/19742554/021314\_1\_online.pdf} \BibitemShut {NoStop}%
\bibitem [{\citenamefont {Christensen}\ \emph {et~al.}(2020)\citenamefont {Christensen}, \citenamefont {Hucul}, \citenamefont {Campbell},\ and\ \citenamefont {Hudson}}]{Christensen:2020dik}%
  \BibitemOpen
  \bibfield  {author} {\bibinfo {author} {\bibfnamefont {J.~E.}\ \bibnamefont {Christensen}}, \bibinfo {author} {\bibfnamefont {D.}~\bibnamefont {Hucul}}, \bibinfo {author} {\bibfnamefont {W.~C.}\ \bibnamefont {Campbell}},\ and\ \bibinfo {author} {\bibfnamefont {E.~R.}\ \bibnamefont {Hudson}},\ }\bibfield  {title} {\bibinfo {title} {High-fidelity manipulation of a qubit enabled by a manufactured nucleus},\ }\href {https://doi.org/10.1038/s41534-020-0265-5} {\bibfield  {journal} {\bibinfo  {journal} {npj Quantum Information}\ }\textbf {\bibinfo {volume} {6}},\ \bibinfo {pages} {35} (\bibinfo {year} {2020})}\BibitemShut {NoStop}%
\bibitem [{\citenamefont {Monroe}\ \emph {et~al.}(2021)\citenamefont {Monroe}, \citenamefont {Campbell}, \citenamefont {Duan}, \citenamefont {Gong}, \citenamefont {Gorshkov}, \citenamefont {Hess}, \citenamefont {Islam}, \citenamefont {Kim}, \citenamefont {Linke}, \citenamefont {Pagano}, \citenamefont {Richerme}, \citenamefont {Senko},\ and\ \citenamefont {Yao}}]{RevModPhys.93.025001}%
  \BibitemOpen
  \bibfield  {author} {\bibinfo {author} {\bibfnamefont {C.}~\bibnamefont {Monroe}}, \bibinfo {author} {\bibfnamefont {W.~C.}\ \bibnamefont {Campbell}}, \bibinfo {author} {\bibfnamefont {L.-M.}\ \bibnamefont {Duan}}, \bibinfo {author} {\bibfnamefont {Z.-X.}\ \bibnamefont {Gong}}, \bibinfo {author} {\bibfnamefont {A.~V.}\ \bibnamefont {Gorshkov}}, \bibinfo {author} {\bibfnamefont {P.~W.}\ \bibnamefont {Hess}}, \bibinfo {author} {\bibfnamefont {R.}~\bibnamefont {Islam}}, \bibinfo {author} {\bibfnamefont {K.}~\bibnamefont {Kim}}, \bibinfo {author} {\bibfnamefont {N.~M.}\ \bibnamefont {Linke}}, \bibinfo {author} {\bibfnamefont {G.}~\bibnamefont {Pagano}}, \bibinfo {author} {\bibfnamefont {P.}~\bibnamefont {Richerme}}, \bibinfo {author} {\bibfnamefont {C.}~\bibnamefont {Senko}},\ and\ \bibinfo {author} {\bibfnamefont {N.~Y.}\ \bibnamefont {Yao}},\ }\bibfield  {title} {\bibinfo {title} {Programmable quantum simulations of spin systems with trapped ions},\ }\href {https://doi.org/10.1103/RevModPhys.93.025001}
  {\bibfield  {journal} {\bibinfo  {journal} {Rev. Mod. Phys.}\ }\textbf {\bibinfo {volume} {93}},\ \bibinfo {pages} {025001} (\bibinfo {year} {2021})}\BibitemShut {NoStop}%
\bibitem [{\citenamefont {Pogorelov}\ \emph {et~al.}(2021)\citenamefont {Pogorelov}, \citenamefont {Feldker}, \citenamefont {Marciniak}, \citenamefont {Postler}, \citenamefont {Jacob}, \citenamefont {Krieglsteiner}, \citenamefont {Podlesnic}, \citenamefont {Meth}, \citenamefont {Negnevitsky}, \citenamefont {Stadler}, \citenamefont {H\"ofer}, \citenamefont {W\"achter}, \citenamefont {Lakhmanskiy}, \citenamefont {Blatt}, \citenamefont {Schindler},\ and\ \citenamefont {Monz}}]{PRXQuantum.2.020343}%
  \BibitemOpen
  \bibfield  {author} {\bibinfo {author} {\bibfnamefont {I.}~\bibnamefont {Pogorelov}}, \bibinfo {author} {\bibfnamefont {T.}~\bibnamefont {Feldker}}, \bibinfo {author} {\bibfnamefont {C.~D.}\ \bibnamefont {Marciniak}}, \bibinfo {author} {\bibfnamefont {L.}~\bibnamefont {Postler}}, \bibinfo {author} {\bibfnamefont {G.}~\bibnamefont {Jacob}}, \bibinfo {author} {\bibfnamefont {O.}~\bibnamefont {Krieglsteiner}}, \bibinfo {author} {\bibfnamefont {V.}~\bibnamefont {Podlesnic}}, \bibinfo {author} {\bibfnamefont {M.}~\bibnamefont {Meth}}, \bibinfo {author} {\bibfnamefont {V.}~\bibnamefont {Negnevitsky}}, \bibinfo {author} {\bibfnamefont {M.}~\bibnamefont {Stadler}}, \bibinfo {author} {\bibfnamefont {B.}~\bibnamefont {H\"ofer}}, \bibinfo {author} {\bibfnamefont {C.}~\bibnamefont {W\"achter}}, \bibinfo {author} {\bibfnamefont {K.}~\bibnamefont {Lakhmanskiy}}, \bibinfo {author} {\bibfnamefont {R.}~\bibnamefont {Blatt}}, \bibinfo {author} {\bibfnamefont {P.}~\bibnamefont {Schindler}},\ and\ \bibinfo {author}
  {\bibfnamefont {T.}~\bibnamefont {Monz}},\ }\bibfield  {title} {\bibinfo {title} {Compact ion-trap quantum computing demonstrator},\ }\href {https://doi.org/10.1103/PRXQuantum.2.020343} {\bibfield  {journal} {\bibinfo  {journal} {PRX Quantum}\ }\textbf {\bibinfo {volume} {2}},\ \bibinfo {pages} {020343} (\bibinfo {year} {2021})}\BibitemShut {NoStop}%
\bibitem [{\citenamefont {Pino}\ \emph {et~al.}(2021)\citenamefont {Pino} \emph {et~al.}}]{Pino:2020mku}%
  \BibitemOpen
  \bibfield  {author} {\bibinfo {author} {\bibfnamefont {J.~M.}\ \bibnamefont {Pino}} \emph {et~al.},\ }\bibfield  {title} {\bibinfo {title} {{Demonstration of the trapped-ion quantum CCD computer architecture}},\ }\href {https://doi.org/10.1038/s41586-021-03318-4} {\bibfield  {journal} {\bibinfo  {journal} {Nature}\ }\textbf {\bibinfo {volume} {592}},\ \bibinfo {pages} {209} (\bibinfo {year} {2021})},\ \Eprint {https://arxiv.org/abs/2003.01293} {arXiv:2003.01293 [quant-ph]} \BibitemShut {NoStop}%
\bibitem [{\citenamefont {Noel}\ \emph {et~al.}(2022)\citenamefont {Noel} \emph {et~al.}}]{Noel:2021hez}%
  \BibitemOpen
  \bibfield  {author} {\bibinfo {author} {\bibfnamefont {C.}~\bibnamefont {Noel}} \emph {et~al.},\ }\bibfield  {title} {\bibinfo {title} {{Measurement-induced quantum phases realized in a trapped-ion quantum computer}},\ }\href {https://doi.org/10.1038/s41567-022-01619-7} {\bibfield  {journal} {\bibinfo  {journal} {Nature Phys.}\ }\textbf {\bibinfo {volume} {18}},\ \bibinfo {pages} {760} (\bibinfo {year} {2022})},\ \Eprint {https://arxiv.org/abs/2106.05881} {arXiv:2106.05881 [quant-ph]} \BibitemShut {NoStop}%
\bibitem [{\citenamefont {Foss-Feig}\ \emph {et~al.}(2024)\citenamefont {Foss-Feig}, \citenamefont {Pagano}, \citenamefont {Potter},\ and\ \citenamefont {Yao}}]{Foss-Feig:2024blk}%
  \BibitemOpen
  \bibfield  {author} {\bibinfo {author} {\bibfnamefont {M.}~\bibnamefont {Foss-Feig}}, \bibinfo {author} {\bibfnamefont {G.}~\bibnamefont {Pagano}}, \bibinfo {author} {\bibfnamefont {A.~C.}\ \bibnamefont {Potter}},\ and\ \bibinfo {author} {\bibfnamefont {N.~Y.}\ \bibnamefont {Yao}},\ }\bibfield  {title} {\bibinfo {title} {{Progress in Trapped-Ion Quantum Simulation}},\ }\href@noop {} {\  (\bibinfo {year} {2024})},\ \Eprint {https://arxiv.org/abs/2409.02990} {arXiv:2409.02990 [quant-ph]} \BibitemShut {NoStop}%
\bibitem [{\citenamefont {Crain}\ \emph {et~al.}(2014)\citenamefont {Crain}, \citenamefont {Mount}, \citenamefont {Baek},\ and\ \citenamefont {Kim}}]{10.1063/1.4900754}%
  \BibitemOpen
  \bibfield  {author} {\bibinfo {author} {\bibfnamefont {S.}~\bibnamefont {Crain}}, \bibinfo {author} {\bibfnamefont {E.}~\bibnamefont {Mount}}, \bibinfo {author} {\bibfnamefont {S.}~\bibnamefont {Baek}},\ and\ \bibinfo {author} {\bibfnamefont {J.}~\bibnamefont {Kim}},\ }\bibfield  {title} {\bibinfo {title} {Individual addressing of trapped 171yb+ ion qubits using a microelectromechanical systems-based beam steering system},\ }\href {https://doi.org/10.1063/1.4900754} {\bibfield  {journal} {\bibinfo  {journal} {Applied Physics Letters}\ }\textbf {\bibinfo {volume} {105}},\ \bibinfo {pages} {181115} (\bibinfo {year} {2014})}\BibitemShut {NoStop}%
\bibitem [{\citenamefont {Shih}\ \emph {et~al.}(2021)\citenamefont {Shih}, \citenamefont {Motlakunta}, \citenamefont {Kotibhaskar}, \citenamefont {Sajjan}, \citenamefont {Habl\"utzel},\ and\ \citenamefont {Islam}}]{Shih:2021fcs}%
  \BibitemOpen
  \bibfield  {author} {\bibinfo {author} {\bibfnamefont {C.-Y.}\ \bibnamefont {Shih}}, \bibinfo {author} {\bibfnamefont {S.}~\bibnamefont {Motlakunta}}, \bibinfo {author} {\bibfnamefont {N.}~\bibnamefont {Kotibhaskar}}, \bibinfo {author} {\bibfnamefont {M.}~\bibnamefont {Sajjan}}, \bibinfo {author} {\bibfnamefont {R.}~\bibnamefont {Habl\"utzel}},\ and\ \bibinfo {author} {\bibfnamefont {R.}~\bibnamefont {Islam}},\ }\bibfield  {title} {\bibinfo {title} {{Reprogrammable and high-precision holographic optical addressing of trapped ions for scalable quantum control}},\ }\href {https://doi.org/10.1038/s41534-021-00396-0} {\bibfield  {journal} {\bibinfo  {journal} {npj Quantum Inf.}\ }\textbf {\bibinfo {volume} {7}},\ \bibinfo {pages} {57} (\bibinfo {year} {2021})}\BibitemShut {NoStop}%
\bibitem [{\citenamefont {Koh}\ \emph {et~al.}(2023)\citenamefont {Koh}, \citenamefont {Sun}, \citenamefont {Motta},\ and\ \citenamefont {Minnich}}]{Koh:2022ajm}%
  \BibitemOpen
  \bibfield  {author} {\bibinfo {author} {\bibfnamefont {J.~M.}\ \bibnamefont {Koh}}, \bibinfo {author} {\bibfnamefont {S.-N.}\ \bibnamefont {Sun}}, \bibinfo {author} {\bibfnamefont {M.}~\bibnamefont {Motta}},\ and\ \bibinfo {author} {\bibfnamefont {A.~J.}\ \bibnamefont {Minnich}},\ }\bibfield  {title} {\bibinfo {title} {{Measurement-induced entanglement phase transition on a superconducting quantum processor with mid-circuit readout}},\ }\href {https://doi.org/10.1038/s41567-023-02076-6} {\bibfield  {journal} {\bibinfo  {journal} {Nature Phys.}\ }\textbf {\bibinfo {volume} {19}},\ \bibinfo {pages} {1314} (\bibinfo {year} {2023})},\ \Eprint {https://arxiv.org/abs/2203.04338} {arXiv:2203.04338 [quant-ph]} \BibitemShut {NoStop}%
\bibitem [{\citenamefont {Ippoliti}\ and\ \citenamefont {Khemani}(2021)}]{PhysRevLett.126.060501}%
  \BibitemOpen
  \bibfield  {author} {\bibinfo {author} {\bibfnamefont {M.}~\bibnamefont {Ippoliti}}\ and\ \bibinfo {author} {\bibfnamefont {V.}~\bibnamefont {Khemani}},\ }\bibfield  {title} {\bibinfo {title} {Postselection-free entanglement dynamics via spacetime duality},\ }\href {https://doi.org/10.1103/PhysRevLett.126.060501} {\bibfield  {journal} {\bibinfo  {journal} {Phys. Rev. Lett.}\ }\textbf {\bibinfo {volume} {126}},\ \bibinfo {pages} {060501} (\bibinfo {year} {2021})}\BibitemShut {NoStop}%
\bibitem [{\citenamefont {Ippoliti}\ \emph {et~al.}(2022)\citenamefont {Ippoliti}, \citenamefont {Rakovszky},\ and\ \citenamefont {Khemani}}]{PhysRevX.12.011045}%
  \BibitemOpen
  \bibfield  {author} {\bibinfo {author} {\bibfnamefont {M.}~\bibnamefont {Ippoliti}}, \bibinfo {author} {\bibfnamefont {T.}~\bibnamefont {Rakovszky}},\ and\ \bibinfo {author} {\bibfnamefont {V.}~\bibnamefont {Khemani}},\ }\bibfield  {title} {\bibinfo {title} {Fractal, logarithmic, and volume-law entangled nonthermal steady states via spacetime duality},\ }\href {https://doi.org/10.1103/PhysRevX.12.011045} {\bibfield  {journal} {\bibinfo  {journal} {Phys. Rev. X}\ }\textbf {\bibinfo {volume} {12}},\ \bibinfo {pages} {011045} (\bibinfo {year} {2022})}\BibitemShut {NoStop}%
\bibitem [{\citenamefont {Lu}\ and\ \citenamefont {Grover}(2021)}]{PRXQuantum.2.040319}%
  \BibitemOpen
  \bibfield  {author} {\bibinfo {author} {\bibfnamefont {T.-C.}\ \bibnamefont {Lu}}\ and\ \bibinfo {author} {\bibfnamefont {T.}~\bibnamefont {Grover}},\ }\bibfield  {title} {\bibinfo {title} {Spacetime duality between localization transitions and measurement-induced transitions},\ }\href {https://doi.org/10.1103/PRXQuantum.2.040319} {\bibfield  {journal} {\bibinfo  {journal} {PRX Quantum}\ }\textbf {\bibinfo {volume} {2}},\ \bibinfo {pages} {040319} (\bibinfo {year} {2021})}\BibitemShut {NoStop}%
\bibitem [{\citenamefont {Potter}\ and\ \citenamefont {Vasseur}(2022)}]{potter2022entanglement}%
  \BibitemOpen
  \bibfield  {author} {\bibinfo {author} {\bibfnamefont {A.~C.}\ \bibnamefont {Potter}}\ and\ \bibinfo {author} {\bibfnamefont {R.}~\bibnamefont {Vasseur}},\ }\bibfield  {title} {\bibinfo {title} {Entanglement dynamics in hybrid quantum circuits},\ }in\ \href@noop {} {\emph {\bibinfo {booktitle} {Entanglement in Spin Chains: From Theory to Quantum Technology Applications}}}\ (\bibinfo  {publisher} {Springer},\ \bibinfo {year} {2022})\ pp.\ \bibinfo {pages} {211--249}\BibitemShut {NoStop}%
\bibitem [{\citenamefont {Dehghani}\ \emph {et~al.}(2023)\citenamefont {Dehghani}, \citenamefont {Lavasani}, \citenamefont {Hafezi},\ and\ \citenamefont {Gullans}}]{Dehghani:2022aia}%
  \BibitemOpen
  \bibfield  {author} {\bibinfo {author} {\bibfnamefont {H.}~\bibnamefont {Dehghani}}, \bibinfo {author} {\bibfnamefont {A.}~\bibnamefont {Lavasani}}, \bibinfo {author} {\bibfnamefont {M.}~\bibnamefont {Hafezi}},\ and\ \bibinfo {author} {\bibfnamefont {M.~J.}\ \bibnamefont {Gullans}},\ }\bibfield  {title} {\bibinfo {title} {{Neural-network decoders for measurement induced phase transitions}},\ }\href {https://doi.org/10.1038/s41467-023-37902-1} {\bibfield  {journal} {\bibinfo  {journal} {Nature Commun.}\ }\textbf {\bibinfo {volume} {14}},\ \bibinfo {pages} {2918} (\bibinfo {year} {2023})},\ \Eprint {https://arxiv.org/abs/2204.10904} {arXiv:2204.10904 [quant-ph]} \BibitemShut {NoStop}%
\bibitem [{\citenamefont {Iouchtchenko}\ \emph {et~al.}(2023)\citenamefont {Iouchtchenko}, \citenamefont {Gonthier}, \citenamefont {Perdomo-Ortiz},\ and\ \citenamefont {Melko}}]{Iouchtchenko:2022php}%
  \BibitemOpen
  \bibfield  {author} {\bibinfo {author} {\bibfnamefont {D.}~\bibnamefont {Iouchtchenko}}, \bibinfo {author} {\bibfnamefont {J.~F.}\ \bibnamefont {Gonthier}}, \bibinfo {author} {\bibfnamefont {A.}~\bibnamefont {Perdomo-Ortiz}},\ and\ \bibinfo {author} {\bibfnamefont {R.~G.}\ \bibnamefont {Melko}},\ }\bibfield  {title} {\bibinfo {title} {{Neural network enhanced measurement efficiency for molecular groundstates}},\ }\href {https://doi.org/10.1088/2632-2153/acb4df} {\bibfield  {journal} {\bibinfo  {journal} {Mach. Learn. Sci. Tech.}\ }\textbf {\bibinfo {volume} {4}},\ \bibinfo {pages} {015016} (\bibinfo {year} {2023})},\ \Eprint {https://arxiv.org/abs/2206.15449} {arXiv:2206.15449 [quant-ph]} \BibitemShut {NoStop}%
\bibitem [{\citenamefont {Czischek}\ \emph {et~al.}(2021{\natexlab{b}})\citenamefont {Czischek}, \citenamefont {Torlai}, \citenamefont {Ray}, \citenamefont {Islam},\ and\ \citenamefont {Melko}}]{Czischek:2021nso}%
  \BibitemOpen
  \bibfield  {author} {\bibinfo {author} {\bibfnamefont {S.}~\bibnamefont {Czischek}}, \bibinfo {author} {\bibfnamefont {G.}~\bibnamefont {Torlai}}, \bibinfo {author} {\bibfnamefont {S.}~\bibnamefont {Ray}}, \bibinfo {author} {\bibfnamefont {R.}~\bibnamefont {Islam}},\ and\ \bibinfo {author} {\bibfnamefont {R.~G.}\ \bibnamefont {Melko}},\ }\bibfield  {title} {\bibinfo {title} {{Simulating a measurement-induced phase transition for trapped ion circuits}},\ }\href {https://doi.org/10.1103/PhysRevA.104.062405} {\bibfield  {journal} {\bibinfo  {journal} {Phys. Rev. A}\ }\textbf {\bibinfo {volume} {104}},\ \bibinfo {pages} {062405} (\bibinfo {year} {2021}{\natexlab{b}})},\ \Eprint {https://arxiv.org/abs/2106.03769} {arXiv:2106.03769 [quant-ph]} \BibitemShut {NoStop}%
\bibitem [{\citenamefont {Gaebler}\ \emph {et~al.}(2016)\citenamefont {Gaebler}, \citenamefont {Tan}, \citenamefont {Lin}, \citenamefont {Wan}, \citenamefont {Bowler}, \citenamefont {Keith}, \citenamefont {Glancy}, \citenamefont {Coakley}, \citenamefont {Knill}, \citenamefont {Leibfried},\ and\ \citenamefont {Wineland}}]{PhysRevLett.117.060505}%
  \BibitemOpen
  \bibfield  {author} {\bibinfo {author} {\bibfnamefont {J.~P.}\ \bibnamefont {Gaebler}}, \bibinfo {author} {\bibfnamefont {T.~R.}\ \bibnamefont {Tan}}, \bibinfo {author} {\bibfnamefont {Y.}~\bibnamefont {Lin}}, \bibinfo {author} {\bibfnamefont {Y.}~\bibnamefont {Wan}}, \bibinfo {author} {\bibfnamefont {R.}~\bibnamefont {Bowler}}, \bibinfo {author} {\bibfnamefont {A.~C.}\ \bibnamefont {Keith}}, \bibinfo {author} {\bibfnamefont {S.}~\bibnamefont {Glancy}}, \bibinfo {author} {\bibfnamefont {K.}~\bibnamefont {Coakley}}, \bibinfo {author} {\bibfnamefont {E.}~\bibnamefont {Knill}}, \bibinfo {author} {\bibfnamefont {D.}~\bibnamefont {Leibfried}},\ and\ \bibinfo {author} {\bibfnamefont {D.~J.}\ \bibnamefont {Wineland}},\ }\bibfield  {title} {\bibinfo {title} {High-fidelity universal gate set for ${^{9}\mathrm{Be}}^{+}$ ion qubits},\ }\href {https://doi.org/10.1103/PhysRevLett.117.060505} {\bibfield  {journal} {\bibinfo  {journal} {Phys. Rev. Lett.}\ }\textbf {\bibinfo {volume} {117}},\ \bibinfo {pages} {060505}
  (\bibinfo {year} {2016})}\BibitemShut {NoStop}%
\bibitem [{\citenamefont {Kong}\ \emph {et~al.}(2024)\citenamefont {Kong}, \citenamefont {Li},\ and\ \citenamefont {Liu}}]{Kong:2024dfn}%
  \BibitemOpen
  \bibfield  {author} {\bibinfo {author} {\bibfnamefont {L.}~\bibnamefont {Kong}}, \bibinfo {author} {\bibfnamefont {Z.}~\bibnamefont {Li}},\ and\ \bibinfo {author} {\bibfnamefont {Z.-W.}\ \bibnamefont {Liu}},\ }\bibfield  {title} {\bibinfo {title} {{Convergence efficiency of quantum gates and circuits}},\ }\href@noop {} {\  (\bibinfo {year} {2024})},\ \Eprint {https://arxiv.org/abs/2411.04898} {arXiv:2411.04898 [quant-ph]} \BibitemShut {NoStop}%
\bibitem [{\citenamefont {HAPPER}(1972)}]{RevModPhys.44.169}%
  \BibitemOpen
  \bibfield  {author} {\bibinfo {author} {\bibfnamefont {W.}~\bibnamefont {HAPPER}},\ }\bibfield  {title} {\bibinfo {title} {Optical pumping},\ }\href {https://doi.org/10.1103/RevModPhys.44.169} {\bibfield  {journal} {\bibinfo  {journal} {Rev. Mod. Phys.}\ }\textbf {\bibinfo {volume} {44}},\ \bibinfo {pages} {169} (\bibinfo {year} {1972})}\BibitemShut {NoStop}%
\bibitem [{\citenamefont {Smith}\ \emph {et~al.}(2024)\citenamefont {Smith}, \citenamefont {Leu}, \citenamefont {Miyanishi}, \citenamefont {Gely},\ and\ \citenamefont {Lucas}}]{Smith:2024gbs}%
  \BibitemOpen
  \bibfield  {author} {\bibinfo {author} {\bibfnamefont {M.~C.}\ \bibnamefont {Smith}}, \bibinfo {author} {\bibfnamefont {A.~D.}\ \bibnamefont {Leu}}, \bibinfo {author} {\bibfnamefont {K.}~\bibnamefont {Miyanishi}}, \bibinfo {author} {\bibfnamefont {M.~F.}\ \bibnamefont {Gely}},\ and\ \bibinfo {author} {\bibfnamefont {D.~M.}\ \bibnamefont {Lucas}},\ }\bibfield  {title} {\bibinfo {title} {{Single-qubit gates with errors at the $10^{-7}$ level}},\ }\href@noop {} {\  (\bibinfo {year} {2024})},\ \Eprint {https://arxiv.org/abs/2412.04421} {arXiv:2412.04421 [quant-ph]} \BibitemShut {NoStop}%
\bibitem [{\citenamefont {Carrasquilla}\ \emph {et~al.}(2019{\natexlab{b}})\citenamefont {Carrasquilla}, \citenamefont {Torlai}, \citenamefont {Melko},\ and\ \citenamefont {Aolita}}]{Carrasquilla_2019}%
  \BibitemOpen
  \bibfield  {author} {\bibinfo {author} {\bibfnamefont {J.}~\bibnamefont {Carrasquilla}}, \bibinfo {author} {\bibfnamefont {G.}~\bibnamefont {Torlai}}, \bibinfo {author} {\bibfnamefont {R.~G.}\ \bibnamefont {Melko}},\ and\ \bibinfo {author} {\bibfnamefont {L.}~\bibnamefont {Aolita}},\ }\bibfield  {title} {\bibinfo {title} {Reconstructing quantum states with generative models},\ }\href {https://doi.org/10.1038/s42256-019-0028-1} {\bibfield  {journal} {\bibinfo  {journal} {Nature Machine Intelligence}\ }\textbf {\bibinfo {volume} {1}},\ \bibinfo {pages} {155–161} (\bibinfo {year} {2019}{\natexlab{b}})}\BibitemShut {NoStop}%
\bibitem [{\citenamefont {Hibat-Allah}\ \emph {et~al.}(2020)\citenamefont {Hibat-Allah}, \citenamefont {Ganahl}, \citenamefont {Hayward}, \citenamefont {Melko},\ and\ \citenamefont {Carrasquilla}}]{hibat2020recurrent}%
  \BibitemOpen
  \bibfield  {author} {\bibinfo {author} {\bibfnamefont {M.}~\bibnamefont {Hibat-Allah}}, \bibinfo {author} {\bibfnamefont {M.}~\bibnamefont {Ganahl}}, \bibinfo {author} {\bibfnamefont {L.~E.}\ \bibnamefont {Hayward}}, \bibinfo {author} {\bibfnamefont {R.~G.}\ \bibnamefont {Melko}},\ and\ \bibinfo {author} {\bibfnamefont {J.}~\bibnamefont {Carrasquilla}},\ }\bibfield  {title} {\bibinfo {title} {Recurrent neural network wave functions},\ }\href {https://doi.org/10.1103/PhysRevResearch.2.023358} {\bibfield  {journal} {\bibinfo  {journal} {Physical Review Research}\ }\textbf {\bibinfo {volume} {2}},\ \bibinfo {pages} {023358} (\bibinfo {year} {2020})}\BibitemShut {NoStop}%
\bibitem [{\citenamefont {Hou}\ \emph {et~al.}(2023)\citenamefont {Hou}, \citenamefont {Li},\ and\ \citenamefont {You}}]{hou2023quantum}%
  \BibitemOpen
  \bibfield  {author} {\bibinfo {author} {\bibfnamefont {W.}~\bibnamefont {Hou}}, \bibinfo {author} {\bibfnamefont {M.}~\bibnamefont {Li}},\ and\ \bibinfo {author} {\bibfnamefont {Y.-Z.}\ \bibnamefont {You}},\ }\bibfield  {title} {\bibinfo {title} {Quantum generative modeling of sequential data with trainable token embedding},\ }\href@noop {} {\bibfield  {journal} {\bibinfo  {journal} {arXiv preprint arXiv:2311.05050}\ } (\bibinfo {year} {2023})}\BibitemShut {NoStop}%
\bibitem [{\citenamefont {Fitzek}\ \emph {et~al.}(2024)\citenamefont {Fitzek}, \citenamefont {Teoh}, \citenamefont {Fung}, \citenamefont {Dagnew}, \citenamefont {Merali}, \citenamefont {Moss}, \citenamefont {MacLellan},\ and\ \citenamefont {Melko}}]{Fitzek:2024onn}%
  \BibitemOpen
  \bibfield  {author} {\bibinfo {author} {\bibfnamefont {D.}~\bibnamefont {Fitzek}}, \bibinfo {author} {\bibfnamefont {Y.~H.}\ \bibnamefont {Teoh}}, \bibinfo {author} {\bibfnamefont {H.~P.}\ \bibnamefont {Fung}}, \bibinfo {author} {\bibfnamefont {G.~A.}\ \bibnamefont {Dagnew}}, \bibinfo {author} {\bibfnamefont {E.}~\bibnamefont {Merali}}, \bibinfo {author} {\bibfnamefont {M.~S.}\ \bibnamefont {Moss}}, \bibinfo {author} {\bibfnamefont {B.}~\bibnamefont {MacLellan}},\ and\ \bibinfo {author} {\bibfnamefont {R.~G.}\ \bibnamefont {Melko}},\ }\bibfield  {title} {\bibinfo {title} {{RydbergGPT}},\ }\href@noop {} {\  (\bibinfo {year} {2024})},\ \Eprint {https://arxiv.org/abs/2405.21052} {arXiv:2405.21052 [quant-ph]} \BibitemShut {NoStop}%
\bibitem [{\citenamefont {Yang}\ \emph {et~al.}(2024)\citenamefont {Yang}, \citenamefont {Soleimanifar}, \citenamefont {Bergamaschi},\ and\ \citenamefont {Preskill}}]{Yang:2024yxu}%
  \BibitemOpen
  \bibfield  {author} {\bibinfo {author} {\bibfnamefont {T.-H.}\ \bibnamefont {Yang}}, \bibinfo {author} {\bibfnamefont {M.}~\bibnamefont {Soleimanifar}}, \bibinfo {author} {\bibfnamefont {T.}~\bibnamefont {Bergamaschi}},\ and\ \bibinfo {author} {\bibfnamefont {J.}~\bibnamefont {Preskill}},\ }\bibfield  {title} {\bibinfo {title} {{When can classical neural networks represent quantum states?}},\ }\href@noop {} {\  (\bibinfo {year} {2024})},\ \Eprint {https://arxiv.org/abs/2410.23152} {arXiv:2410.23152 [quant-ph]} \BibitemShut {NoStop}%
\bibitem [{\citenamefont {Cho}\ \emph {et~al.}(2014)\citenamefont {Cho}, \citenamefont {van Merri{\"e}nboer}, \citenamefont {Bahdanau},\ and\ \citenamefont {Bengio}}]{cho-etal-2014-properties}%
  \BibitemOpen
  \bibfield  {author} {\bibinfo {author} {\bibfnamefont {K.}~\bibnamefont {Cho}}, \bibinfo {author} {\bibfnamefont {B.}~\bibnamefont {van Merri{\"e}nboer}}, \bibinfo {author} {\bibfnamefont {D.}~\bibnamefont {Bahdanau}},\ and\ \bibinfo {author} {\bibfnamefont {Y.}~\bibnamefont {Bengio}},\ }\bibfield  {title} {\bibinfo {title} {On the properties of neural machine translation: Encoder{--}decoder approaches},\ }in\ \href {https://doi.org/10.3115/v1/W14-4012} {\emph {\bibinfo {booktitle} {Proceedings of {SSST}-8, Eighth Workshop on Syntax, Semantics and Structure in Statistical Translation}}},\ \bibinfo {editor} {edited by\ \bibinfo {editor} {\bibfnamefont {D.}~\bibnamefont {Wu}}, \bibinfo {editor} {\bibfnamefont {M.}~\bibnamefont {Carpuat}}, \bibinfo {editor} {\bibfnamefont {X.}~\bibnamefont {Carreras}},\ and\ \bibinfo {editor} {\bibfnamefont {E.~M.}\ \bibnamefont {Vecchi}}}\ (\bibinfo  {publisher} {Association for Computational Linguistics},\ \bibinfo {address} {Doha, Qatar},\ \bibinfo {year} {2014})\ pp.\
  \bibinfo {pages} {103--111}\BibitemShut {NoStop}%
\bibitem [{\citenamefont {Vaswani}(2017)}]{vaswani2017attention}%
  \BibitemOpen
  \bibfield  {author} {\bibinfo {author} {\bibfnamefont {A.}~\bibnamefont {Vaswani}},\ }\bibfield  {title} {\bibinfo {title} {Attention is all you need},\ }\href@noop {} {\bibfield  {journal} {\bibinfo  {journal} {Advances in Neural Information Processing Systems}\ } (\bibinfo {year} {2017})}\BibitemShut {NoStop}%
\bibitem [{\citenamefont {Zhang}\ and\ \citenamefont {Di~Ventra}(2023)}]{PhysRevB.107.075147}%
  \BibitemOpen
  \bibfield  {author} {\bibinfo {author} {\bibfnamefont {Y.-H.}\ \bibnamefont {Zhang}}\ and\ \bibinfo {author} {\bibfnamefont {M.}~\bibnamefont {Di~Ventra}},\ }\bibfield  {title} {\bibinfo {title} {Transformer quantum state: A multipurpose model for quantum many-body problems},\ }\href {https://doi.org/10.1103/PhysRevB.107.075147} {\bibfield  {journal} {\bibinfo  {journal} {Phys. Rev. B}\ }\textbf {\bibinfo {volume} {107}},\ \bibinfo {pages} {075147} (\bibinfo {year} {2023})}\BibitemShut {NoStop}%
\bibitem [{\citenamefont {Lange}\ \emph {et~al.}(2024)\citenamefont {Lange}, \citenamefont {Bornet}, \citenamefont {Emperauger}, \citenamefont {Chen}, \citenamefont {Lahaye}, \citenamefont {Kienle}, \citenamefont {Browaeys},\ and\ \citenamefont {Bohrdt}}]{lange2024transformer}%
  \BibitemOpen
  \bibfield  {author} {\bibinfo {author} {\bibfnamefont {H.}~\bibnamefont {Lange}}, \bibinfo {author} {\bibfnamefont {G.}~\bibnamefont {Bornet}}, \bibinfo {author} {\bibfnamefont {G.}~\bibnamefont {Emperauger}}, \bibinfo {author} {\bibfnamefont {C.}~\bibnamefont {Chen}}, \bibinfo {author} {\bibfnamefont {T.}~\bibnamefont {Lahaye}}, \bibinfo {author} {\bibfnamefont {S.}~\bibnamefont {Kienle}}, \bibinfo {author} {\bibfnamefont {A.}~\bibnamefont {Browaeys}},\ and\ \bibinfo {author} {\bibfnamefont {A.}~\bibnamefont {Bohrdt}},\ }\bibfield  {title} {\bibinfo {title} {Transformer neural networks and quantum simulators: a hybrid approach for simulating strongly correlated systems},\ }\href@noop {} {\bibfield  {journal} {\bibinfo  {journal} {arXiv preprint arXiv:2406.00091}\ } (\bibinfo {year} {2024})}\BibitemShut {NoStop}%
\bibitem [{\citenamefont {Zhang}\ and\ \citenamefont {You}(2024)}]{Zhang_2024}%
  \BibitemOpen
  \bibfield  {author} {\bibinfo {author} {\bibfnamefont {Z.}~\bibnamefont {Zhang}}\ and\ \bibinfo {author} {\bibfnamefont {Y.-Z.}\ \bibnamefont {You}},\ }\bibfield  {title} {\bibinfo {title} {Observing schrödinger’s cat with artificial intelligence: emergent classicality from information bottleneck},\ }\href {https://doi.org/10.1088/2632-2153/ad3330} {\bibfield  {journal} {\bibinfo  {journal} {Machine Learning: Science and Technology}\ }\textbf {\bibinfo {volume} {5}},\ \bibinfo {pages} {015051} (\bibinfo {year} {2024})}\BibitemShut {NoStop}%
\bibitem [{\citenamefont {Yao}\ and\ \citenamefont {You}(2024)}]{yao2024shadowgpt}%
  \BibitemOpen
  \bibfield  {author} {\bibinfo {author} {\bibfnamefont {J.}~\bibnamefont {Yao}}\ and\ \bibinfo {author} {\bibfnamefont {Y.-Z.}\ \bibnamefont {You}},\ }\bibfield  {title} {\bibinfo {title} {Shadowgpt: Learning to solve quantum many-body problems from randomized measurements},\ }\href@noop {} {\bibfield  {journal} {\bibinfo  {journal} {arXiv preprint arXiv:2411.03285}\ } (\bibinfo {year} {2024})}\BibitemShut {NoStop}%
\bibitem [{\citenamefont {Wang}\ \emph {et~al.}(2024)\citenamefont {Wang}, \citenamefont {Li}, \citenamefont {Chen}, \citenamefont {Cheng}, \citenamefont {Liu},\ and\ \citenamefont {Chen}}]{Wang:2024pdt}%
  \BibitemOpen
  \bibfield  {author} {\bibinfo {author} {\bibfnamefont {H.}~\bibnamefont {Wang}}, \bibinfo {author} {\bibfnamefont {P.}~\bibnamefont {Li}}, \bibinfo {author} {\bibfnamefont {M.}~\bibnamefont {Chen}}, \bibinfo {author} {\bibfnamefont {J.}~\bibnamefont {Cheng}}, \bibinfo {author} {\bibfnamefont {J.}~\bibnamefont {Liu}},\ and\ \bibinfo {author} {\bibfnamefont {T.}~\bibnamefont {Chen}},\ }\bibfield  {title} {\bibinfo {title} {{GroverGPT: A Large Language Model with 8 Billion Parameters for Quantum Searching}},\ }\href@noop {} {\  (\bibinfo {year} {2024})},\ \Eprint {https://arxiv.org/abs/2501.00135} {arXiv:2501.00135 [quant-ph]} \BibitemShut {NoStop}%
\bibitem [{\citenamefont {Paszke}\ \emph {et~al.}(2019)\citenamefont {Paszke}, \citenamefont {Gross}, \citenamefont {Massa}, \citenamefont {Lerer}, \citenamefont {Bradbury}, \citenamefont {Chanan}, \citenamefont {Killeen}, \citenamefont {Lin}, \citenamefont {Gimelshein}, \citenamefont {Antiga} \emph {et~al.}}]{paszke2019pytorch}%
  \BibitemOpen
  \bibfield  {author} {\bibinfo {author} {\bibfnamefont {A.}~\bibnamefont {Paszke}}, \bibinfo {author} {\bibfnamefont {S.}~\bibnamefont {Gross}}, \bibinfo {author} {\bibfnamefont {F.}~\bibnamefont {Massa}}, \bibinfo {author} {\bibfnamefont {A.}~\bibnamefont {Lerer}}, \bibinfo {author} {\bibfnamefont {J.}~\bibnamefont {Bradbury}}, \bibinfo {author} {\bibfnamefont {G.}~\bibnamefont {Chanan}}, \bibinfo {author} {\bibfnamefont {T.}~\bibnamefont {Killeen}}, \bibinfo {author} {\bibfnamefont {Z.}~\bibnamefont {Lin}}, \bibinfo {author} {\bibfnamefont {N.}~\bibnamefont {Gimelshein}}, \bibinfo {author} {\bibfnamefont {L.}~\bibnamefont {Antiga}}, \emph {et~al.},\ }\bibfield  {title} {\bibinfo {title} {Pytorch: An imperative style, high-performance deep learning library},\ }\href@noop {} {\bibfield  {journal} {\bibinfo  {journal} {Advances in neural information processing systems}\ }\textbf {\bibinfo {volume} {32}} (\bibinfo {year} {2019})}\BibitemShut {NoStop}%
\end{thebibliography}%
\end{document}